\pdfoutput=1

\documentclass[
aps,
prd,
10pt,
twocolumn,
superscriptaddress,
preprintnumbers,
nofootinbib
]{revtex4-2}
\usepackage{physics}
\usepackage{xcolor}

\newif\ifdraft
\drafttrue
\usepackage[T1]{fontenc}

\usepackage{braket}
\usepackage{titlesec}
\usepackage{
	amsmath,
	amssymb,
	amsfonts,
	amsthm,
	mathtools,
	mathrsfs,
	bm,
	stmaryrd,
	dsfont,
	leftidx,
	accents,
	slashed,
	cancel,
	empheq
}

\usepackage{graphicx}
\graphicspath{{Images/}}

\usepackage{tikz}
\usepackage{tikz-3dplot}
\usepackage{pgfplots}

\pgfplotsset{compat=1.18}

\usetikzlibrary{
	arrows.meta,
	backgrounds,
	calc,
	decorations.markings,
	decorations.pathmorphing,
	decorations.pathreplacing,
	intersections,
	patterns,
	patterns.meta,
	shapes
}

\usepgfplotslibrary{fillbetween}

\tikzset{
	snakeit/.style={
		decorate,
		decoration=snake
	},
	zigzag/.style={
		decorate,
		decoration=zigzag
	}
}

\usepackage{array}
\usepackage{multirow}
\usepackage{tabularx}
\usepackage{diagbox}

\usepackage{enumitem}

\usepackage{subfig}

\usepackage{comment}
\usepackage{ifthen}
\usepackage{import}
\usepackage{wrapfig}
\usepackage[percent]{overpic}
\usepackage[most]{tcolorbox}
\usepackage{mdframed}
\usepackage{pifont}

\usepackage[dvipsnames]{xcolor}

\definecolor{rosy}{RGB}{230,235,252}
\definecolor{myframetitle}{RGB}{90,89,170}
\definecolor{myblocktitle}{RGB}{140,185,249}
\definecolor{mytitle}{RGB}{10,80,26}

\definecolor{darkgreen}{RGB}{27,130,45}
\definecolor{darkblue}{rgb}{0,0,0.3}
\definecolor{darkred}{rgb}{0.7,0,0}

\definecolor{lightgray}{RGB}{220,220,220}
\definecolor{darkpurple}{RGB}{108,0,217}

\definecolor{pink}{RGB}{190,20,100}
\definecolor{orang}{RGB}{193,63,0}
\definecolor{green}{RGB}{11,98,17}
\definecolor{darkpink}{RGB}{153,0,76}
\definecolor{bluegreen}{RGB}{0,102,102}
\definecolor{greenlagan}{RGB}{0,102,0}
\definecolor{redgreen}{RGB}{102,102,0}
\definecolor{Redgreen}{RGB}{153,76,0}

\definecolor{vividviolet}{rgb}{0.62,0.0,1.0}
\definecolor{amaranth}{rgb}{0.9,0.17,0.31}
\definecolor{palatinateblue}{rgb}{0.15,0.23,0.89}
\definecolor{brightpink}{rgb}{1.0,0.0,0.5}
\definecolor{cornflowerblue}{rgb}{0.39,0.58,0.93}
\definecolor{deepcarminepink}{rgb}{0.94,0.19,0.22}
\definecolor{radicalred}{rgb}{1.0,0.21,0.37}

\usepackage[
colorlinks=true,
linktoc=all,
linkcolor=palatinateblue,
citecolor=brightpink,
urlcolor=amaranth,
pagebackref=true
]{hyperref}

\renewcommand*{\backref}[1]{}
\renewcommand*{\backrefalt}[4]{%
	\ifcase #1%
	\or
	(Cited in p.~#2)%
	\else
	(Cited in pp.~#2)%
	\fi
}

\usepackage{etoolbox}

\makeatletter
\apptocmd{\NAT@sort@cites}
{\expandafter\Hy@backout\expandafter{\NAT@cite@list}}
{}
{\PackageError{backref-fix}{Could not patch natbib}{}}
\makeatother

\ifdraft

\newcommand\hnote[1]{%
	\textcolor{blue}{\bfseries [HA:\,#1]}%
}

\newcommand\anote[1]{%
	\textcolor{magenta}{\bfseries [AL:\,#1]}%
}

\newenvironment{redinsert}
{\begingroup\color{red}}
{\endgroup}

\else

\newcommand\hnote[1]{}
\newcommand\anote[1]{}

\hypersetup{hidelinks}

\fi

\newtheorem{theorem}{Theorem}

\newtheorem{definition}[theorem]{Definition}

\newcommand{\dif}{\mathrm{d}}

\tcbset{
	highlight math style={
		left=2mm,
		right=2mm,
		top=2mm,
		bottom=2mm
	}
}

\newcommand{\be}{\begin{equation}}
	\newcommand{\ee}{\end{equation}}

\begin{document}
	%%%%%%%%%%%%%%%%%%%%%%%%%%%%%%%%%%%%%%%%%%%%%%%%%%
	
	\title{Fluid/Gravity Correspondence in 3d Flat Gravity with Non--Compact Boundaries}
	
	%%%%%%%%%%%%%%%%%%%%%%%%%%%%%%%%%%%%%%%%%%%%%%%%%%
	% Authors
	%%%%%%%%%%%%%%%%%%%%%%%%%%%%%%%%%%%%%%%%%%%%%%%%%%
	
	\author{H.~Adami}
	\email{hadami@simis.cn}
	
	\affiliation{
		Center for Mathematics and Interdisciplinary Sciences,
		Fudan University,
		Shanghai 200433,
		China
	}
	
	\affiliation{
		Shanghai Institute for Mathematics and Interdisciplinary Sciences (SIMIS),
		Shanghai 200433,
		China
	}
	
	\author{,~R.~Esp\'indola}
	\email{ricardo.esro1@gmail.com}
	
	\affiliation{Institute for Advanced Study, Tsinghua University, Beijing 100084, China}
	
	\author{,~K. ~Lara}
	\email{klarad@cecs.cl}
	
	\affiliation{Centro de Estudios Cient\'{\i}ficos (CECs), Av. Arturo Prat 514, Valdivia, Chile.}
	\affiliation{Facultad de Ingenier\'ia, Universidad San Sebasti\'an, sede Valdivia, General Lagos 1163, Valdivia
		5110693, Chile}
	
	\author{A.~Latifi}
	\email{latifi@qut.ac.ir}
	
	\affiliation{Department of Mechanics, Qom University of Technology, Qom, Iran }

	\begin{abstract}
		We study \emph{3d} flat gravity on a restricted phase space
		with a non-compact spatial boundary. The charge algebra splits into two commuting
		chiral current algebras, which a fluid/gravity dictionary maps to the stress tensor
		of a two-dimensional conformal fluid, with energy bounded below by the momentum.
		On the line, finiteness of the charges selects integrable, square-integrable
		currents, and the global data are carried by the scattering variables of a
		Schr\"odinger operator rather than by a monodromy. We analyze Heisenberg-like and
		Dirichlet-like boundary conditions; the latter give independently propagating
		chiral kinks associated with the reflectionless P\"oschl--Teller operator. Both
		admit first-law relations. We outline the quantization as a Weyl
		$C^*$-algebra with quasi-free states. Retaining the radial dependence, the
		quasi-local stress tensor obeys an exact, non-perturbative flow of square-root
		$T\bar T$ type, which integrates to a finite boost in the radial position. 
	\end{abstract}
	
	\maketitle
	
	\tableofcontents
	
	%%%%%%%%%%%%%%%%%%%%%%%%%%%%%%%%%%%%%%%%%%%%%%%%%%%%%%%
	\section{Introduction}
	%%%%%%%%%%%%%%%%%%%%%%%%%%%%%%%%%%%%%%%%%%%%%%%%%%%%%%%

	Understanding gravitational dynamics on finite-distance timelike or null
	surfaces, rather than only at asymptotic infinity, has become a central theme in
	holography. In the AdS$_3$/CFT$_2$ context, the $T\bar T$ deformation provides a
	field-theoretic description of gravity with a finite radial cutoff
	\cite{Hartman:2018tkw}, a picture that has since been refined through the
	relation to timelike Liouville theory \cite{Allameh:2025gsa}. Analogous
	constructions have been developed for flat space, where the finite-cutoff
	problem is tied to the structure of asymptotically flat gravity
	\cite{Banihashemi:2024yye}. Related work has explored how asymptotically flat
	spacetimes can be related to finite causal diamonds \cite{Ciambelli:2025fbo},
	and how Rindler fluids arise from gravitational shockwaves \cite{Bak:2024kzk}.
	The latter fits into the broader fluid/gravity and membrane-paradigm tradition
	\cite{Damour:1979wya,Thorne:1986iy,Hubeny:2010ry,Bredberg:2011jq,Penna:2015gza,Arenas-Henriquez:2023hur}.
	
	In three dimensions, flat-space holography is governed by the BMS$_3$ asymptotic
	symmetry algebra. Its most general fall-off \footnote{with fixed chemical potentials at the asymptotic region.} and enhanced symmetry
	algebras have been studied extensively
	\cite{Grumiller:2017sjh,Detournay:2016sfv}. Together with the thermodynamics and
	holography of flat cosmological horizons \cite{Barnich:2012xq,Bagchi:2012xr},
	this has motivated a search for the dual field theory
	\cite{Bagchi:2016bcd,Afshar:2019axx,Basak:2023otu,Despontin:2025dog}. Further
	developments include modular Hamiltonians and entanglement in flat holography
	\cite{Apolo:2020qjm,Caceres:2026yvs}, geometric actions \cite{Merbis:2019wgk},
	BMS particles \cite{Oblak:2016eij}, and holographic realizations of Carrollian
	free fields, including higher-spin and infinite-spin generalizations
	\cite{DArcy:2026hgl,Buchbinder:2023qog}.

	Integrable systems with BMS$_3$ Poisson structures has also attracted attention, where the ``flat KdV'' hierarchy emerge as the asymptotic dynamics of locally flat spacetimes~\cite{Fuentealba:2017omf} (see, e.g., further works \cite{Oblak:2020bkg} and \cite{Vitel:2026ssa}.)

	Integrability provides a parallel thread running through two-dimensional
	conformal field theory and gravity. Generalized Gibbs ensembles built from KdV
	charges, and their holographic duals, offer a window into the statistics of
	black-hole microstates \cite{Dymarsky:2020tjh,Maloney:2018yrz,Chen:2024ysb},
	while the integrable structure of CFT is encoded in quantum KdV theory
	\cite{Bazhanov:1994ft}. 
	
	%Integrable black-hole dynamics and classical integrability in AdS$_3$ asymptotic symmetries have been analysed in \cite{Cardenas:2025qka,Cardenas:2025qqi}, 
	On AdS$_3$, the geometrization of non-trivial profiles of boundary KdV and Dym movers has been analyzed in \cite{Cardenas:2025qka} and \cite{Pino:2025crn}, respectively (a review on the topic can be found in \cite{Cardenas:2025qka}), and integrable-system techniques have
	also appeared in the study of inhomogeneous quenches and Floquet CFT
	\cite{Jiang:2024hgt}, three-dimensional C-metrics \cite{Tian:2024mew}, and
	holographic entanglement with conformal boundary conditions
	\cite{Caceres:2026yvs}. The phase-space structures underlying these results are
	naturally formulated using Virasoro coadjoint orbits
	\cite{Balog:1997zz,Navarro-Salas:1999ejl,Raeymaekers:2014kea}, together with
	Poisson-Lie and symplectic structures \cite{Balog:1999zm,Alekseev:1993qs}. They
	also relate to the algebraic and operator-algebraic study of chiral conformal
	field theories and their modular properties
	\cite{Bakalov:1995um,Morinelli:2021nsx,weiner2007conformal,Adamo:2025aqk}. In AdS/CFT, modular parallel transport relates boundary Berry curvature to bulk symplectic forms associated with entanglement wedges \cite{deBoer:2021zlm, Czech:2023zmq}.
	Related geometric insights into conformal field theory and entropy
	\cite{Martinec:1998wm}, as well as soft Heisenberg hair on black holes
	\cite{Afshar:2016wfy}, provide further context.
	
	Null surfaces offer a complementary arena in which these ideas can be made
	precise. The phase space, thermodynamics, and stress-tensor structure of null
	boundaries, together with their soft-hair interpretation and near-horizon
	dynamics, have been developed in a number of works
	\cite{Adami:2024rkr,Adami:2023wbe,Adami:2021nnf,Adami:2021kvx,Adami:2020amw,Grumiller:2018scv,Grumiller:2019tyl,Grumiller:2019fmp},
	building on the construction of Brown--York charges and symmetries at null
	boundaries \cite{Chandrasekaran:2021hxc,Chandrasekaran:2018aop}. A complementary
	approach based on quantum null geometry, the null Raychaudhuri equation, and the
	full phase space of gravity on null hypersurfaces
	\cite{Ciambelli:2026xdn,Ciambelli:2024swv,Ciambelli:2023mir} has been developed
	alongside Carrollian hydrodynamics on stretched horizons
	\cite{Freidel:2022vjq,Freidel:2022bai}. At null infinity \cite{Bondi:1962px,Sachs:1962wk,Sachs:1962zza,Barnich:2006av,Barnich:2010eb,Barnich:2012rz,Barnich:2014kra,Barnich:2015sca}, Carrollian techniques
	underlie the celestial-holography programme \cite{Donnay:2022aba}, the
	derivation of gravitational flux-balance laws \cite{Fiorucci:2025twa,Fiorucci:2026nyg,Arenas-Henriquez:2025rpt}; for comprehensive reviews see
	\cite{Ruzziconi:2026bix,Hartong:2025jpp,Hartong:2026bvh}.
	
	Holography at a finite radial cutoff provides a geometric realization of
	Wilsonian renormalization group flow, relating bulk radial evolution to the
	dynamics of the dual quantum field theory away from its ultraviolet fixed point
	\cite{deBoer:1999xf,Bianchi:2001kw,Skenderis:2002wp,Papadimitriou:2004ap,Heemskerk:2010hk,Faulkner:2010jy,Adami:2025pqr}.
	A central result in this direction is the proposal that a finite cutoff in
	asymptotically AdS$_3$ gravity corresponds to a $T\bar T$ deformation of the
	dual two-dimensional conformal field theory \cite{McGough:2016lol}, where
	$T\bar T$ is a solvable irrelevant deformation generated by the determinant of
	the stress tensor \cite{Zamolodchikov:2004ce,Smirnov:2016lqw,Cavaglia:2016oda}.
	Its exact finite-volume energy spectrum and retained integrable properties make
	this deformation a particularly tractable framework for studying holographic
	renormalization group flows. The correspondence has been investigated through
	energy spectra, thermodynamics, transport, correlation functions, and the radial
	evolution of the Brown--York stress tensor
	\cite{Hartman:2018tkw,Aharony:2018bad,Taylor:2018xcy,Kraus:2018xrn,Guica:2019nzm},
	and extended to theories with additional conserved currents, higher-spin fields,
	supersymmetry, non-relativistic symmetries, and related irrelevant deformations
	\cite{Guica:2017lia,Chakraborty:2018vja,Apolo:2018qpq}.
	
	Within the broader setting of AdS$_3$/CFT$_2$
	\cite{Brown:1986nw,Coussaert:1995zp,Achucarro:1987vz,Witten:1988hc,Strominger:1997eq,Maldacena:1997re,Witten:1998qj,Gubser:1998bc},
	substantial evidence supports interpreting a finite radial cutoff in the bulk as
	a deformation of the dual two-dimensional conformal field theory by
	$T\bar T$-type operators, including the $\sqrt{T\bar T}$ operator
	\cite{Rodriguez:2021tcz,Tempo:2022ndz,Babaei-Aghbolagh:2022uij,Babaei-Aghbolagh:2022leo,Conti:2022egv,Ferko:2022cix,Babaei-Aghbolagh:2024hti,Ebert:2023tih,Banerjee:2026qyc}.
	These deformations preserve integrability while modifying the ultraviolet
	behavior of the theory, leading to characteristic finite-cutoff effects such as
	a deformed energy spectrum, altered thermodynamics, and state-dependent
	effective geometries
	\cite{Guica:2019nzm,Hartman:2018tkw,Aharony:2018bad,Taylor:2018xcy}. In the
	holographic description, the radial location of the cutoff hypersurface is
	related to the energy scale of the deformed theory.
	
	However, existing analyses have primarily established this correspondence
	through the finite-volume spectrum or through trace relations for the
	renormalized quasi-local stress tensor
	\cite{McGough:2016lol,Hartman:2018tkw,Guica:2019nzm}. These quantities do not fully
	determine the radial evolution of local observables, and they do not generally
	yield a closed local evolution equation for the stress tensor alone, with the
	radial evolution instead commonly derived from a Fefferman--Graham expansion.
	An exact, non-perturbative description of the quasi-local Brown--York stress
	tensor at finite radius would therefore provide a more direct characterization
	of the holographic dictionary beyond the asymptotic boundary, and a
	complementary formulation of finite-cutoff holography in which the radial
	dynamics of the boundary stress tensor is made explicit.

	The paper is organized as follows. Section~\ref{sec:CS} reviews the formulation of three-dimensional flat gravity as an $\mathfrak{isl}(2,\mathbb{R})$ Chern--Simons theory and introduces the associated renormalized symplectic structure. In Sections \ref{Section: Restricted phase space}, \ref{Section: Constructing the metric} and \ref{Section: Generic solution}, we subsequently specify the restricted phase space, reconstruct the spacetime metric, and solve the equations of motion in terms of scalar potentials, respectively. The Killing symmetries and zero modes are analyzed in Sec.~\ref{sec:killing-symmetries}, followed by a study of the charge algebra and its decomposition into two chiral sectors in Sec.~\ref{sec:Diagonal-phase-space}. Section~\ref{sec:fluid-gravity} develops the fluid/gravity correspondence, while Sec.~\ref{sec:Boundary-metric} examines the boundary geometry, the role of Weyl rescaling, and the resulting energy--momentum bound. The renormalized symplectic form on a non-compact slice is investigated in Sec.~\ref{sec:symplectic}. The spectral analysis of the chiral currents, including the associated Schr\"odinger operators and transport matrix, is presented in Sec.~\ref{app:schrodinger-transport-matrix}. We then study two classes of boundary conditions: Heisenberg-like conditions in Sec.~\ref{Heisenberg-BC} and Dirichlet-like conditions in Sec.~\ref{Dirichlet-BC}, the latter leading to a P\"oschl--Teller system. Section~\ref{sec:quantization} addresses quantization on the real line using the Weyl $C^*$-algebra, the GNS construction, and a continuum mode decomposition. In Sec.~\ref{sqrtTTbar radial flow}, we restore the radial dependence to derive the exact square-root radial flow equation for the quasi-local stress tensor and integrate it in terms of a finite $SO(1,1)$ boost. Finally, Sec.~\ref{sec:conc} summarizes the results, discusses their interpretation within the membrane paradigm, and outlines the limitations of the present construction and possible directions for further investigation.

	%%%%%%%%%%%%%%%%%%%%%%%%%%%%%%%%%%%%%%%%%%%%%%%%%%%%%%%
	\section{\emph{3d} flat gravity as a Chern--Simons theory}\label{sec:CS}
	%%%%%%%%%%%%%%%%%%%%%%%%%%%%%%%%%%%%%%%%%%%%%%%%%%%%%%%
	
	Pure Einstein gravity in three spacetime dimensions admits an equivalent first-order formulation as a Chern--Simons (CS) gauge theory. In the absence of a cosmological constant, the relevant gauge algebra is $\mathfrak{isl}(2,\mathbb{R})$, the Lie algebra of the three-dimensional Poincar\'e group.
	
	The CS action for a $\mathfrak{isl}(2,\mathbb{R})$-valued connection $A$ on a three-manifold $\mathcal{M}$ is
	\begin{equation}\label{eq:CS-action}
		\begin{split}
			S_{\text{\tiny CS}}[A]
			={}&
			\frac{k}{4\pi}
			\int_{\mathcal{M}}
			\Big\langle
			A \wedge \dif A
			+\frac{2}{3} A \wedge A \wedge A
			\Big\rangle 
			\\
			& +  \int_{\partial\mathcal{M}}
			\text{L}_{\text{\tiny bdy}}[A] \, ,
		\end{split}
	\end{equation}
	where $k$ is the CS level, $\langle\cdot,\cdot\rangle$ is a nondegenerate invariant bilinear form on the gauge algebra, and $\text{L}_{\text{\tiny bdy}}$ is a boundary term chosen to ensure a well-defined variational principle for the boundary conditions of interest. We adopt units in which the CS level is fixed to $k=2\pi$. This choice is equivalent to setting $8\pi G=1$, with $G$ denoting Newton's gravitational constant.
	
	A convenient basis of $\mathfrak{isl}(2,\mathbb{R})$ consists of the Lorentz generators $L_m$ and translation generators $M_m$, with $m\in\{-1,0,1\}$. Their commutation relations are
	\begin{equation}\label{eq:isl2alg}
		\begin{split}
			&  [L_m,L_n] = f_{mn}{}^p \, L_{p}\,  , \qquad
			[M_m,L_n] = f_{mn}{}^p \, M_{p}\,  , \\
			& [M_m,M_n] =0\,  ,
		\end{split}
	\end{equation}
	where the structure constants are
	\begin{equation}
		f_{mn}{}^p:=(m-n)\,\delta^p_{m+n}\, .
	\end{equation}
	The generators $L_m$ span an $\mathfrak{sl}(2,\mathbb{R})$ subalgebra, while the $M_m$ form an abelian ideal transforming in the adjoint representation of $\mathfrak{sl}(2,\mathbb{R})$.
	
	The gauge algebra admits a unique nondegenerate invariant bilinear form (up to an overall normalization), which may be chosen as
	\begin{equation}\label{eq:kappa}
		\kappa_{mn}
		=
		\frac{1}{2}\langle L_m,M_n\rangle
		=
		\begin{pmatrix}
			0 & 0 & -1 \\
			0 & \frac12 & 0 \\
			-1 & 0 & 0
		\end{pmatrix}.
	\end{equation}
	All remaining pairings vanish, $\langle L_m,L_n\rangle=0 $ and $ \langle M_m,M_n\rangle=0 $.
	The existence of this nondegenerate bilinear form is essential for the CS formulation, as it provides the gauge-invariant pairing entering the action \eqref{eq:CS-action}.
	
	Varying the CS Lagrangian gives
	\begin{equation}
		\delta \text{L}_{\text{\tiny CS}}[A]
		=  \langle    F \wedge \delta A \rangle +  \dif  \Theta_{\text{\tiny CS}}[\delta A;A] ,
	\end{equation}
	where $F=\dif A+A\wedge A$ is the curvature two-form associated with the gauge connection $A$. The total derivative term defines the symplectic potential,
	\begin{equation}
		\Theta_{\text{\tiny CS}}[\delta A;A]
		= -\frac{1}{2}\, \langle   A \wedge \delta A \rangle +  \delta \text{L}_{\text{\tiny bdy}}[A] ,
	\end{equation}
	where the second contribution originates from the variation of the boundary Lagrangian. The corresponding Euler--Lagrange equations are obtained by requiring the bulk variation to vanish for arbitrary $\delta A$, leading to $F=0$. Thus, classical solutions are flat gauge connections. The dynamics is therefore entirely encoded in the global properties of the gauge field and, in the presence of boundaries, in the boundary degrees of freedom determined by the choice of boundary conditions and the boundary term $\text{L}_{\text{\tiny bdy}}$.

	To characterize the phase space, it is convenient to adopt the standard radial gauge decomposition of the CS connection,
	\begin{equation}\label{eq:radialgauge}
		A =  b^{-1}(r) \left(  \dif  + a  \right)  b(r)\, ,
	\end{equation}
	where the group-valued function $b(r)$ depends exclusively on the radial coordinate. The auxiliary connection $a$ is defined solely on the constant-$r$ hypersurfaces and can be written as
	\begin{equation}\label{gauge-connection-parameterized}
		a = a_t \dif  t + a_x \dif  x \, .
	\end{equation}
	This choice of gauge completely fixes the radial dependence of the bulk connection, rendering it pure gauge. Throughout the following analysis, the dynamical fields are therefore identified with the components of $a$, while the radial group element $b(r)$ serves only to reconstruct the full bulk connection without introducing additional physical degrees of freedom.
	
	Residual gauge transformations that preserve both the radial gauge condition \eqref{eq:radialgauge} and the parametrization of the gauge connection given in \eqref{gauge-connection-parameterized} are generated by infinitesimal gauge parameters of the form
	\begin{equation}\label{eq:trans-law}
		\delta_\epsilon a
		= \dif \epsilon + [a,\epsilon],
	\end{equation}
	The gauge parameter $\epsilon$ is an arbitrary $\mathfrak{isl}(2,\mathbb{R})$ valued function of the boundary coordinates. 
	
	The divergences in the on-shell action can be removed by supplementing the bulk CS action with the boundary term
	\begin{equation}
		\text{L}_{\text{\tiny bdy}}[A] =\frac{1}{2}\, \langle   \dif b\, b^{-1} \wedge a \rangle \, .
	\end{equation}
	With this choice, the variational principle is well defined under the prescribed boundary conditions, and the resulting action is finite on the constant-$r$ hypersurface. Accordingly, the renormalized (regularized) symplectic potential is obtained by combining the bulk contribution with the variation of the boundary term. Restricting to the boundary hypersurface $\mathcal{B}$ at fixed $r$, one finds
	\begin{equation}
		\boldsymbol{\Theta}_{\text{\tiny ren.}}[\delta A;A]
		= -\frac{1}{2} \int_{\mathcal{B}} \langle   a \wedge \delta a \rangle \, .
	\end{equation}
	
	The renormalized symplectic form is defined as the field-space exterior derivative of the renormalized symplectic potential,
	\begin{equation}
		\boldsymbol{\Omega}_{\text{\tiny ren.}}[\delta A, \delta A;A]
		= \delta  \boldsymbol{\Theta}_{\text{\tiny ren.}}[\delta A;A]\, ,
	\end{equation}
	and is therefore closed in field space. Its contraction with an infinitesimal gauge transformation generated by $\epsilon$ takes the form
	\begin{equation}
		\boldsymbol{\Omega}_{\text{\tiny ren.}}[\delta A, \delta_\epsilon A;A]
		= \int_{\mathcal{B}} \dif \langle   \epsilon \, \delta a\rangle ,
	\end{equation}
	which implies that the corresponding gauge transformation gives rise to a boundary contribution to the symplectic structure. Within the covariant phase space formalism
	\cite{Lee:1990nz,Iyer:1994ys,Wald:1999wa}, this boundary term determines the variation of the associated surface charge. Accordingly, the variation of the surface charge corresponding to the gauge parameter $\epsilon$ is
	\begin{equation}
		\delta Q(\epsilon) =  \int_{\Sigma} \langle \epsilon\,\delta a \rangle \, ,
		\label{eq:charge-variation}
	\end{equation}
	where $\Sigma\subset\mathcal{B}$ denotes a constant-time spatial section of the boundary. Since $\mathcal{B}$ is two-dimensional, $\Sigma$ is a one-dimensional codimension-two surface of the bulk spacetime.
	
	The above equation also leads to a balance law for the charge variation. Integrating the boundary contribution over the portion of $\mathcal{B}$ bounded by the constant-time slices at $t=t_i$ and $t=t_f$, and keeping track of possible contributions from the spatial endpoints, one obtains
	\begin{equation}
		\begin{split}
			\delta Q(\epsilon)\Big|_{t_f}
			-
			\delta Q(\epsilon)\Big|_{t_i}
			=& \ \boldsymbol{\Omega}_{\text{\tiny ren.}}[\delta A, \delta_\epsilon A;A]\, \\
			&-  \int_{t_i}^{t_f} \langle \epsilon\,\delta a \rangle \Big|_{x=-\infty}^{x=+\infty}\, .
			\label{eq:charge-balance}
		\end{split}
	\end{equation}
	where $x=\pm\infty $ denote boundaries of $\Sigma$.
	Equation \eqref{eq:charge-balance} shows that the difference between the charge variations evaluated on the two time slices is controlled by the renormalized symplectic flux through the intervening portion of the boundary, together with a possible contribution from its spatial endpoints. Therefore, the associated surface charge is conserved only when the right-hand side of \eqref{eq:charge-balance} vanishes. In particular, conservation follows when the relevant symplectic flux vanishes and no flux is exchanged through the spatial endpoints. The latter condition is automatically satisfied for a periodic spatial section, while for an interval it requires boundary conditions that eliminate the corresponding endpoint contribution. More generally, \eqref{eq:charge-balance} should be understood as a charge balance relation: non-vanishing symplectic or endpoint flux accounts for the variation of the charge between the two constant-time slices.

	%%%%%%%%%%%%%%%%%%%%%%%%%%%%%%%%%%%%%%%%%%%%%%
	\section{Restricted phase space}\label{Section: Restricted phase space}
	%%%%%%%%%%%%%%%%%%%%%%%%%%%%%%%%%%%%%%%%%%%%%%
	
	A particularly simple and physically relevant sector of the phase space is obtained by truncating the $\pm1$ components of the state-dependent functions and chemical potentials. 
	
	The reduced connection is restricted to the Cartan subalgebra of
	$\mathfrak{isl}(2,\mathbb{R})$ and therefore contains only the zero components of
	the Lorentz and translation generators. It takes the form
	\begin{equation}
		a= \mathcal{A}^{\text{\tiny L}}\, L_0 + \mathcal{A}^{\text{\tiny M}} \, M_0 \, .
	\end{equation}
	The associated one-forms are parametrized as \cite{Afshar:2016wfy,Grumiller:2017sjh}
	\begin{subequations}
		\begin{align}
			\mathcal{A}^{\text{\tiny L}} ={} & 2\, \mu \,\dif t + \frac{1}{2} \, \mathcal{L} \,\dif x \, , \\
			\mathcal{A}^{\text{\tiny M}} ={} & 2\, \nu \,\dif t + \frac{1}{2} \, \mathcal{P} \,\dif x \, .
		\end{align}
	\end{subequations}
	Here, $\mathcal{A}^{\text{\tiny L}}$ and $\mathcal{A}^{\text{\tiny M}}$
	encode the residual boundary data associated with the Lorentz and translation
	sectors, respectively.

	This restriction defines a consistent subsector of the full phase space since the equations of motion preserve the above conditions. Indeed, upon substituting the truncation into the equations of motion, one finds
	\begin{equation}\label{eq:EOM-RPS}
		\partial_t \mathcal{L} =4 \, \partial_x \mu \, , \qquad \partial_t\mathcal{P}  =4 \,\partial_x \nu \, .
	\end{equation}
	Hence, the dynamics is entirely governed by the remaining fields and their associated chemical potentials.
	
	Consistency further requires that the residual gauge transformations preserve the restricted phase space. The residual gauge parameter
	\begin{equation}
		\epsilon
		=2\,\zeta \,L_0 +2\, \xi \, M_0 \,,
	\end{equation}
	is completely characterized by the components $\zeta$ and $\xi$, which generate transformations preserving the restricted boundary conditions. Imposing this condition on the transformation laws \eqref{eq:trans-law} yields
	\begin{subequations}\label{transformation-laws-RPS}
		\begin{align}
			& \delta_\epsilon \mathcal{L} = 4 \,\partial_x \zeta  \, , & & 
			\delta_\epsilon \mathcal{P}  = 4 \,\partial_x\xi  \, , \\
			&\delta_\epsilon \mu =   \partial_t \zeta  \, , & & 
			\delta_\epsilon \nu = \partial_t\xi \, .
		\end{align}
	\end{subequations}
	
	Specializing the general expressions for the renormalized symplectic structure to this sector, the renormalized symplectic potential and symplectic form reduce to
	\begin{subequations}
		\begin{align}
			\boldsymbol{\Theta}_{\text{\tiny ren.}}
			={} & -\int_{\mathcal{B}} \left( \mu \, \delta \mathcal{P} +\nu \, \delta \mathcal{L} \right) \dif  t   \dif  x  \nonumber\\
			&+\frac{1}{2}\,\delta \! \int_{\mathcal{B}} \left( \mu  \, \mathcal{P} +\nu \,   \mathcal{L} \right) \dif  t   \dif  x \,  , \label{renormalized-symplectic-potential}\\
			\boldsymbol{\Omega}_{\text{\tiny ren.}}
			={} & - \int_{\mathcal{B}} \left( \delta \mu \curlywedge \delta \mathcal{P} +\delta \nu \curlywedge  \delta \mathcal{L} \right) \dif  t   \dif  x \,  .\label{renormalized-symplectic-form}
		\end{align}
	\end{subequations}
	The resulting symplectic form is manifestly finite and depends only on the canonically conjugate pairs $(\mu,\mathcal{P})$ and $(\nu,\mathcal{L})$. Consequently, the restricted phase space inherits a well-defined renormalized symplectic structure from the full theory.
	
	The variation of the canonical surface charge associated with the residual gauge parameter $\epsilon=(\xi,\zeta)$ is given by
	\begin{equation}\label{C-V-RPS-new}
		\delta Q(\epsilon)
		=
		\int_{\Sigma}
		\left(
		\xi\,\delta\mathcal{L}
		+
		\zeta\,\delta\mathcal{P}
		\right)
		\dif x \, .
	\end{equation}
	
	%%%%%%%%%%%%%%%%%%%%%%%%%%%%%%%%%%%%%%%%%%%%%%
	\section{Constructing the metric}\label{Section: Constructing the metric}
	%%%%%%%%%%%%%%%%%%%%%%%%%%%%%%%%%%%%%%%%%%%%%%

	The spacetime metric is obtained from the translational component of the CS connection by employing the invariant bilinear form of
	$\mathfrak{isl}(2,\mathbb{R})$. In particular, the metric is reconstructed
	according to
	\begin{equation}\label{line-element}
		\dif s^2 :=\frac{1}{2} \,\kappa^{mn} \, \langle A L_m \rangle\langle A L_n \rangle .
	\end{equation}
	To introduce the radial direction, we choose the group element
	\begin{equation}
		b(r)= \exp\!\left[\frac{r}{2}\, ( M_{+1}- M_{-1})\right] .
	\end{equation}
	This choice fixes the radial dependence of the full CS connection through the usual radial decomposition. Evaluating the bilinear form with the
	resulting connection gives the metric
	\begin{equation}\label{eq:metric-generic}
		\dif s^2=  \dif r^2   -r^2\,  (\mathcal{A}^{\text{\tiny L}})^2+  (\mathcal{A}^{\text{\tiny M}})^2   \, .
	\end{equation}
	The resulting line element exhibits a natural radial foliation, in which the
	constant-$r$ hypersurfaces are determined by the boundary one-forms
	$\mathcal{A}^{\text{\tiny L}}$ and $\mathcal{A}^{\text{\tiny M}}$. The entire
	radial dependence is generated by the group element $b(r)$, while the remaining
	degrees of freedom are encoded in the reduced connection.
	
	For the metric \eqref{eq:metric-generic}, the corresponding invariant volume element is determined by
	\begin{equation}
		\sqrt{-g}=r \, |\mu \, \mathcal{P}- \nu \,\mathcal{L}| \, .
	\end{equation}
	The absolute value guarantees that the volume element is non-negative, as required by its geometric definition. Throughout this work, we therefore restrict attention to the non-degenerate sector, $\mu \,\mathcal{P}-\nu \,\mathcal{L}\neq 0$ for which the metric is invertible and the corresponding geometry is well defined.

	%%%%%%%%%%%%%%%%%%%%%%%%%%%%%%%%%%%%%%%%%%%%%%%%%%%%%%%%%%%%
	\section{Generic solution}\label{Section: Generic solution}
	%%%%%%%%%%%%%%%%%%%%%%%%%%%%%%%%%%%%%%%%%%%%%%%%%%%%%%%%%%%%
	
	The equations of motion~\eqref{eq:EOM-RPS} constitute a set of first-order constraints. Their local solution can be conveniently parametrized in terms of two scalar potentials, $T(t,x)$ and $X(t,x)$, associated with the two independent sectors of the theory. In this parametrization, the charge densities take the form of total spatial derivatives,
	\begin{equation}
		\label{eq:L-M-to-T-X}
		\mathcal{L}(t,x)
		=
		2\,\partial_x T(t,x) \, , \qquad \mathcal{P}(t,x)
		=
		2\,\partial_x X(t,x)\, ,
	\end{equation}
	which makes the constrained nature of the reduced phase space manifest.
	
	Substituting~\eqref{eq:L-M-to-T-X} into the equations of motion determines the chemical potentials in terms of the time evolution of the scalar potentials, up to the usual freedom associated with the choice of chiral representatives. A convenient parametrization that solves~\eqref{eq:EOM-RPS} identically is
	\begin{equation}
		\label{eq:mu-nu-to-T-X}
		\mu(t,x)
		=
		\frac{1}{2} \partial_t T(t,x)\, , \qquad \nu(t,x)
		=
		\frac{1}{2} \partial_t X(t,x) \, .
	\end{equation}
	Therefore, the functions $T$ and $X$ provide a complete set of local coordinates on the classical solution space within the considered sector.
	
	Employing the reconstruction relation between the CS variables and the boundary data, the corresponding spacetime metric takes the form
	\begin{equation}\label{eq:Rindler-metric}
		\dif s^2 = \dif r^2 -r^2 \dif T^2 +\dif X^2 \, ,
	\end{equation}
	This geometry is locally flat and corresponds to Minkowski spacetime written in Rindler-type coordinates. Consequently, it solves the vacuum Einstein equations with vanishing cosmological constant. The arbitrary functions encoded in $T$ and $X$ characterize different representatives of the reduced phase space. In particular, configurations related by improper boundary diffeomorphisms correspond to physically distinct solutions and represent boundary excitations, commonly referred to as boundary gravitons.

	For the metric~\eqref{eq:Rindler-metric}, the determinant in the coordinates $(r,T,X)$ is $\sqrt{-g}=r$. The apparent singular behavior at $r=0$ is the standard coordinate degeneracy associated with the Rindler horizon rather than a physical curvature singularity. Likewise, the asymptotic region $r\rightarrow+\infty$ does not correspond to a singular locus of the spacetime, but rather to the boundary of the chosen coordinate patch.

	The transformation laws \eqref{transformation-laws-RPS} induce simple variations for the scalar potentials parametrizing the solution space. In particular, one finds
	\begin{equation}
		\delta_\epsilon  T(t,x) = 2 \, \zeta  \, , \qquad 
		\delta_\epsilon X  = 2 \, \xi  \, .
	\end{equation}
	Thus, the functions $T$ and $X$ transform directly under the residual symmetries through shifts determined by the corresponding symmetry parameters. This confirms that the scalar potentials provide a natural description of the reduced phase space, with the action of the residual transformations realized as translations in field space.

	%%%%%%%%%%%%%%%%%%%%%%%%%%%%%%%%%%%%%%%%%%%%%%%%%%%%%%%%%%%%
	\section{Killing symmetries and zero modes}\label{sec:killing-symmetries}
	%%%%%%%%%%%%%%%%%%%%%%%%%%%%%%%%%%%%%%%%%%%%%%%%%%%%%%%%%%%%
	
	Transformation laws \eqref{transformation-laws-RPS} admit a distinguished subclass of transformations generated by field-independent parameters. These correspond to rigid symmetries of the phase space and leave the dynamical configuration invariant. Specifically, by restricting the gauge parameters to constant values, say $\xi
	=1$ and/or $\zeta
	=1$, the induced variations of the dynamical fields vanish. Hence these transformations generate isometries of the phase space and may be regarded as Killing symmetries.
	
	The charges associated with these rigid symmetries follow from the general surface-charge expression evaluated on constant parameters, 
	\begin{subequations}\label{eq:zero-mode-charges}
		\begin{align}
			& \mathrm{P}_0 = \int_{\Sigma} \mathcal{P}\, 
			\dif x =  2 \left[  X(t,+\infty)-X(t,-\infty)\right]\, , \\
			& \mathrm{L}_0 = \int_{\Sigma}
			\mathcal{L}\,\dif x
			= 2 \left[  T(t,+\infty)-T(t,-\infty)\right]  \, , 
		\end{align}
	\end{subequations}
	denote the zero-mode charges. These quantities represent the global information carried by the field configuration.
	
	Moreover, the quantities $\mathrm{P}_0$ and $\mathrm{L}_0$ characterize disconnected sectors of the phase space and act as labels of superselection sectors. Their physical interpretation depends on the global properties of the target space of $T$ and $X$. For non-compact target spaces, they assume continuous values and distinguish inequivalent asymptotic configurations.

	Using \eqref{eq:charge-balance}, the contribution of the symplectic form vanishes when the associated symmetry generator is Killing. Consequently, the corresponding zero-mode charges are conserved, provided that the time derivatives of the symmetry parameters satisfy the matching conditions at the two endpoints of the spatial interval,
	\begin{subequations}
		\begin{align}
			& \partial_t T(t,+\infty)=\partial_t T(t,-\infty)\, ,\\
			& \partial_t X(t,+\infty)=\partial_t X(t,-\infty)\, ,
		\end{align}
	\end{subequations}
	or equivalently
	\begin{equation}\label{conservation-condition}
		\mu(t,+\infty)=\mu(t,-\infty)\, ,\qquad \nu(t,+\infty)=\nu(t,-\infty) \, .
	\end{equation}
	These conditions ensure that no net boundary contribution to the charge balance arises from the endpoints. Therefore, although the local charge density may in general evolve nontrivially, its zero mode remains time-independent under the above endpoint conditions.
	
	We postulate that the zero modes are invariant under the infinitesimal symmetry transformations, $\delta_\epsilon \mathrm{P}_0=\delta_\epsilon \mathrm{L}_0=0$, while they remain nontrivial directions in field space, $\delta \mathrm{P}_0\neq 0$ and $\delta \mathrm{L}_0\neq 0$, the transformation laws impose restrictions on the allowed symmetry parameters. In particular, the invariance of the zero modes requires the corresponding symmetry generators to satisfy the following boundary conditions along the spatial direction
	\begin{equation}\label{BC-symmetry-generators}
		\xi(t,+\infty)=\xi(t,-\infty)\, ,\qquad \zeta(t,+\infty)=\zeta(t,-\infty) \, .
	\end{equation}
	These conditions ensure that the symmetry transformations preserve the zero modes without restricting their independent variations over the phase space.

	%%%%%%%%%%%%%%%%%%%%%%%%%%%%%%%%%%%%%%%%%%%%%%%%%%%%%%%%%%%%
	\section{Charge Algebra}
	%%%%%%%%%%%%%%%%%%%%%%%%%%%%%%%%%%%%%%%%%%%%%%%%%%%%%%%%%%%%
	
	When the gauge parameters $\xi$ and $\zeta$ are treated as field-independent test functions, the surface charge variation \eqref{C-V-RPS-new} is manifestly integrable in field space. The corresponding finite surface charge therefore takes the form
	\begin{equation}\label{C-RPS-new}
		Q(\epsilon)
		=
		\int_{\Sigma}
		\left(
		\xi\,\mathcal{L}
		+
		\zeta\,\mathcal{P}
		\right)
		\dif x \, .
	\end{equation}
	The integrability of the charges follows directly from the absence of field-dependent contributions in the symmetry parameters, so that no nontrivial integration path in field space is required.
	
	The algebra of surface charges is defined through the covariant phase space prescription
	\begin{equation}\label{charge-algebra-def}
		\left\{
		Q(\epsilon_1),
		Q(\epsilon_2)
		\right\}
		:=
		\delta_{\epsilon_2}Q(\epsilon_1)\, .
	\end{equation}
	Employing the transformation laws \eqref{transformation-laws-RPS} and assuming that the test functions satisfy \eqref{BC-symmetry-generators}, so that all boundary contributions arising from integrations by parts can be discarded, one obtains the equal-time on-shell Poisson brackets
	\begin{subequations}\label{charge-algebra}
		\begin{align}
			& \left\{
			\mathcal{L}(t,x),
			\mathcal{P}(t,y)
			\right\}
			=
			4\,
			\partial_x
			\delta(x-y)\, ,\\
			& \left\{
			\mathcal{L}(t,x),
			\mathcal{L}(t,y)
			\right\}
			=
			0 \, ,
			\quad
			\left\{
			\mathcal{P}(t,x),
			\mathcal{P}(t,y)
			\right\}
			=
			0 \, ,
		\end{align}
	\end{subequations}
	where $\delta(x-y)$ stands for the Dirac delta distribution.\footnote{ For a point $y\in \Sigma$, the Dirac delta distribution $\delta(x-y)$ on the interval $\Sigma$ is defined through its action on a sufficiently regular function $f(x)$ as
		\begin{equation*}
			\int_{-\infty}^{+\infty} d x\,
			\delta(x-y)f(x)
			=
			f(y).
		\end{equation*}
	} These brackets define an infinite-dimensional Heisenberg algebra. In particular, both $\mathcal{L}$ and $\mathcal{P}$ separately generate abelian current algebras, while their mixed bracket contains the only nontrivial central extension proportional to the CS level $k$.

	%%%%%%%%%%%%%%%%%%%%%%%%%%%%%%%%%%%%%%%%%%%%%%%
	\section{Diagonal phase space: change of basis}\label{sec:Diagonal-phase-space}
	%%%%%%%%%%%%%%%%%%%%%%%%%%%%%%%%%%%%%%%%%%%%%%%
	
	It is convenient to introduce variables that diagonalize the boundary dynamics. To this end, we define
	\begin{equation}\label{dictionary}
		\mathcal{J}_\pm:=\frac{1}{2} \left( \mathcal{P} \pm \mathcal{L}\right)\, , \qquad \mu_\pm:= \mu \pm \nu \, .
	\end{equation}
	In terms of these combinations, the equations of motion decouple into two independent chiral sectors,
	\begin{equation}\label{EOM-chiral}
		\partial_t \mathcal{J}_\pm = \pm 2 \, \partial_x \mu_\pm \, .
	\end{equation}
	The fields $\mathcal{J}_+$ and $\mathcal{J}_-$ therefore propagate independently, with opposite chiralities, while $\mu_\pm$ play the role of their corresponding chemical potentials.
	
	The renormalized symplectic potential and symplectic form also take a diagonal form,
	\begin{subequations} \begin{align} &\boldsymbol{\Theta}_{\text{\tiny ren.}} = -\int_{\mathcal{B}} \left( \mu_+ \, \delta \mathcal{J}_+ +\mu_- \, \delta \mathcal{J}_- \right) \dif t \dif x \nonumber \\
			& \qquad+\frac{1}{2}\,\delta \! \int_{\mathcal{B}} \left( \mu_+ \, \mathcal{J}_+ +\mu_- \mathcal{J}_- \right) \dif t \dif x \, ,\\
			&\boldsymbol{\Omega}_{\text{\tiny ren.}} = - \int_{\mathcal{B}} \left( \delta \mu_+ \curlywedge \delta \mathcal{J}_+ +\delta \mu_- \curlywedge \delta \mathcal{J}_- \right) \dif t \dif x \, . \end{align} \end{subequations}
	
	The absence of mixed $(+,-)$ contributions shows that the phase space factorizes into two mutually independent sectors. Consequently, each chiral sector constitutes an independent canonical subsystem with conjugate variables $(\mu_\pm,\mathcal{J}_\pm)$.
	
	Equation \eqref{dictionary} establishes the correspondence between the variables of the two descriptions and provides the essential dictionary that enables the direct translation of results obtained in asymptotically AdS$_3$ spacetimes to the asymptotically locally flat case. In particular, this map allows us to extend the analysis of \cite{Adami:2026jmo} and formulate the fluid/gravity correspondence for asymptotically locally flat geometries. Throughout the remainder of this section, we systematically employ the dictionary \eqref{dictionary} to reinterpret the gravitational dynamics in terms of the corresponding fluid variables and derive the associated fluid equations and constitutive relations in the flat-space setting.
	
	It is useful to introduce the combinations $\eta_\pm=\zeta\pm\xi$ in terms of which the residual gauge transformations \eqref{transformation-laws-RPS} take a manifestly diagonal form,
	\begin{equation}\label{eq:gaugetransf}
		\delta_\epsilon \mathcal{J}_\pm
		=
		\pm
		2
		\partial_x \eta_\pm ,
		\qquad
		\delta_\epsilon \mu_\pm
		=
		\partial_t \eta_\pm .
	\end{equation}
	Thus, the two sectors labeled by $\pm$ transform independently. This observation motivates the parametrization
	\begin{equation}
		\label{eq:J-to-Phi}
		\mathcal{J}_\pm
		=
		2\,\partial_x \Phi_\pm \, , \qquad \mu_\pm
		=
		\pm \partial_t \Phi_\pm \, ,
	\end{equation}
	where $\Phi_\pm=(X\pm T)/2$. This representation makes explicit the local potential description of the diagonal currents.
	
	For the representation in terms of the asymptotic values of $\Phi_\pm$ to be meaningful, the latter must also admit finite limits as $x\to\pm\infty$, with their difference determined by the integral of the current. Thus, the $L^1(\mathbb{R})$ condition provides a natural sufficient criterion for the existence of finite zero-mode charges on the noncompact spatial slice. It is worth emphasizing, however, that finiteness of the displayed integral is weaker than absolute integrability in general: conditionally convergent integrals may exist even when a function does not belong to $L^1(\mathbb{R})$. Accordingly, the statement $\mathcal{J}_\pm\in L^1(\mathbb{R})$ should be understood as the natural and robust assumption ensuring that the zero-mode charges are well defined without relying on cancellations at spatial infinity.
	
	The charge algebra \eqref{charge-algebra} determines the corresponding equal-time Poisson brackets of the boundary currents on shell. The two sectors form mutually commuting current algebras, with opposite signs for their central terms:
	\begin{subequations}\label{charge-algebra-chiral}
		\begin{align}
			& \left\{\mathcal J_{\pm}(t,x),\mathcal J_{\pm}(t,y)\right\}
			=\pm 2\, \partial_x\delta(x-y)\, ,\\
			& \left\{\mathcal J_{+}(t,x),\mathcal J_{-}(t,y)\right\}
			=
			0 \, .
		\end{align}
	\end{subequations}
	Thus, the $+$ and $-$ sectors decouple at the level of the equal-time Poisson algebra, while each sector carries a field-independent central extension whose sign is fixed by the corresponding chirality.

	%%%%%%%%%%%%%%%%%%%%%%%%%%%%%%%%%%%%%%%%%%%%%%%%%%%%%%%%%%%%
	\section{Fluid/Gravity Correspondence}\label{sec:fluid-gravity}
	%%%%%%%%%%%%%%%%%%%%%%%%%%%%%%%%%%%%%%%%%%%%%%%%%%%%%%%%%%%%
	
	In this section we establish the fluid/gravity correspondence for the
	restricted phase space constructed in the previous sections. The basic
	idea is to interpret the induced geometry on the timelike boundary as
	the background spacetime on which an effective relativistic fluid
	propagates, while the gravitational phase-space variables determine the
	fluid stress tensor. The correspondence is implemented through the
	dictionary introduced in Section~\ref{sec:Diagonal-phase-space}, which
	relates the gravitational variables to the hydrodynamic data.
	
	Let $(\mathcal{N},\gamma)$ be a two-dimensional Lorentzian manifold
	equipped with the metric
	\begin{equation}\label{Boundary-metric-fixed}
		\dif \gamma^2
		=
		-N^2 \dif t^2
		+ \left(\dif x + N^x \dif t\right)^2 .
	\end{equation}
	Guided by the holographic dictionary developed in
	Section~\ref{sec:Diagonal-phase-space}, we identify the lapse and shift
	functions as \footnote{In terms of chiral quantities, we have 
		\begin{equation*}
			N = 
			\frac{\mu_+}{\mathcal{J}_+}
			+
			\frac{\mu_-}{\mathcal{J}_-} \, ,
			\qquad
			N^x = 
			\frac{\mu_+}{\mathcal{J}_+}
			-
			\frac{\mu_-}{\mathcal{J}_-}
			\, .
		\end{equation*}
	}
	\begin{equation}
		N =
		\frac{4 \left(\mu\,\mathcal{P}-\nu\,\mathcal{L}\right)}
		{\mathcal{P}^2-\mathcal{L}^2} \, , \qquad
		N^x = 
		\frac{4 \left(\nu\,\mathcal{P}-\mu\,\mathcal{L}\right)}
		{\mathcal{P}^2-\mathcal{L}^2} \, .
	\end{equation}

	These expressions encode the boundary chemical potentials in terms of
	the ADM variables of the boundary geometry and provide the bridge
	between the gravitational and hydrodynamic descriptions.
	
	An orthonormal coframe adapted to the ADM foliation is given by
	\begin{equation}
		u_a \dif x^a = -N \dif t \, ,
		\qquad
		s_a \dif x^a = \dif x + N^x \dif t \, ,
	\end{equation}
	with dual frame
	\begin{equation}
		u^a \partial_a
		=
		N^{-1}
		\left(
		\partial_t-N^x\partial_x
		\right),
		\qquad
		s^a\partial_a
		=
		\partial_x \, .
	\end{equation}
	By construction, these vectors satisfy
	\begin{equation}
		u_a u^a=-1,
		\qquad
		s_a s^a=1,
		\qquad
		u_a s^a=0 \, ,
	\end{equation}
	so that $\{u^a,s^a\}$ forms an orthonormal basis on tangent bundle on
	$\mathcal{N}$. The future-directed timelike vector $u^a$ is naturally
	identified with the fluid velocity, whereas $s^a$ spans the unique
	spatial direction orthogonal to the flow.
	
	In two spacetime dimensions, any symmetric rank-two tensor can be
	expanded uniquely in this orthonormal basis. Consequently, the most
	general energy-momentum tensor takes the form
	\begin{equation}\label{BY-EMT}
		\mathcal{T}^{ab}
		=
		\rho\,u^a u^b
		+
		p\,s^a s^b
		+
		2J\,u^{(a}s^{b)} \, ,
	\end{equation}
	where $\rho$, $p$, and $J$ are functions on $\mathcal{N}$.
	Respectively, they represent the local energy density, pressure, and
	energy flux (or, equivalently, momentum density) measured in the local
	rest frame defined by $u^a$. The trace of the energy-momentum tensor is
	\begin{equation}
		\mathcal{T}
		:=
		\gamma_{ab}\mathcal{T}^{ab}
		=
		-\rho+p \, .
	\end{equation}
	Unlike fluids in higher dimensions, a $(1+1)$-dimensional relativistic
	fluid possesses neither shear nor independent anisotropic stresses.
	Accordingly, the three functions $\rho$, $p$, and $J$ completely
	characterize the stress tensor.
	
	The fluid dynamics is determined by the covariant conservation law
	\begin{equation}\label{Conservation-equations}
		\nabla_b\mathcal{T}^{ab}=0 \, ,
	\end{equation}
	where $\nabla_a$ denotes the metric--connection compatible with the
	boundary metric $\gamma_{ab}$. Projecting
	\eqref{Conservation-equations} along $u^a$ and $s^a$ yields the two
	independent hydrodynamic conservation equations governing the evolution
	of the energy and momentum densities.
	
	The constitutive relations are fixed by combining the conservation
	equations with the variational principle encoded in the renormalized
	symplectic potential,
	\begin{equation}\label{SP-Hydro}
		\boldsymbol{\Theta}_{\text{\tiny ren.}}
		=
		-\frac{1}{2}
		\int_{\mathcal{N}}
		\sqrt{-\gamma}\,
		\mathcal{T}^{ab}\,
		\delta\gamma_{ab}\,
		\dif t \dif x \, ,
	\end{equation}
	which identifies $\mathcal{T}^{ab}$ as the response of the renormalized
	gravitational action to variations of the boundary metric. Requiring
	simultaneously that the stress tensor satisfy
	\eqref{Conservation-equations} uniquely determines the hydrodynamic
	variables in terms of the gravitational phase-space fields,
	\begin{equation}\label{eq:fluid-density-relations}
		\rho =
		\frac{1}{8}
		\left(
		\mathcal{P}^2+\mathcal{L}^2
		\right),
		\quad
		J=
		-\frac{1}{4}\,
		\mathcal{L}\,\mathcal{P}
		\, , \quad
		p=\rho \, .
	\end{equation}
	The resulting equation of state, $p=\rho$, immediately implies that the stress tensor is traceless, $\mathcal{T}=0$, which is consistent with the conformal nature of the effective
	two-dimensional fluid. Therefore, the entire hydrodynamic description is
	completely determined by the two gravitational phase-space variables
	$\mathcal{L}$ and $\mathcal{P}$. This establishes an explicit correspondence between the gravitational
	phase space and the dynamics of a relativistic conformal fluid living on
	the timelike boundary.
	
	We can now define the total energy and total momentum of the fluid by integrating the corresponding local densities over the spatial hypersurface $\Sigma$. Explicitly,
	\begin{subequations}\label{QLEM}
		\begin{align}
			\mathscr{E}
			& :=
			\int_{\Sigma}
			\rho\,\dif x
			=
			\frac{1}{8}
			\int_{\Sigma}
			\left(
			\mathcal{P}^2+\mathcal{L}^2
			\right)\dif x \, , \label{QLE}
			\\
			\mathscr{J}
			& :=
			\int_{\Sigma}
			J\,\dif x
			=
			-\frac{1}{4}
			\int_{\Sigma}
			\mathcal{L}\,\mathcal{P}\,\dif x\, . \label{QLM}
		\end{align}
	\end{subequations}
	The total energy is manifestly non-negative, whereas the total momentum is not sign-definite. It is useful to express these quantities in terms of the chiral combinations $\mathcal{J}_\pm$. Defining
	\begin{equation}\label{P-M-H}
		\mathscr{H}_\pm = \frac{1}{4}
		\int_{\Sigma}
		\mathcal{J}_\pm^2\, \dif x \, ,
	\end{equation}
	one obtains 
	\begin{equation}\label{E-J-H-p-m}
		\mathscr{E}= \mathscr{H}_+ +\mathscr{H}_- \, , \qquad \mathscr{J}= -\mathscr{H}_+ +\mathscr{H}_- \, .
	\end{equation}
	Thus, the total energy is the sum of two non-negative chiral contributions, while the momentum measures their difference, with the relative sign fixed by the definitions of the chiral sectors.
	
	The conservation of the total energy and momentum consequently requires the corresponding chiral contributions to be conserved separately. Using the equations of motion, their time evolution can be written as
	\begin{equation}\label{C-E-M-C}
		\partial_t \mathscr{H}_\pm = \pm 
		\int_{\Sigma}
		\mathcal{J}_\pm\,\partial_x \mu_\pm \dif x \, ,
	\end{equation}
	Hence, the conservation of $\mathscr{H}_\pm$ is controlled by the spatial behavior of the chiral currents and their associated chemical potentials. In particular, whether these quantities define conserved charges on a non-compact spatial hypersurface depends not only on the local equations of motion but also on the admissible boundary behavior of the fields at the endpoints of $\Sigma$.
	
	The finiteness of the energy and momentum requires the chiral currents to satisfy both integrability and square-integrability conditions. In particular, one must have not only $\mathcal{J}_{\pm}\in L^1(\mathbb{R})$, but also $\mathcal{J}_{\pm}\in L^2(\mathbb{R})$. Hence, the allowed phase space is naturally restricted to configurations for which the chiral currents belong to $L^1(\mathbb{R})\cap L^2(\mathbb{R})$.

	%%%%%%%%%%%%%%%%%%%%%%%%%%%%%%%%%%%%%%%%%%%%%%%%%%%%%%%%%%%%
	\section{On boundary metric}\label{sec:Boundary-metric}
	%%%%%%%%%%%%%%%%%%%%%%%%%%%%%%%%%%%%%%%%%%%%%%%%%%%%%%%%%%%%
	
	In this section we clarify the geometrical and physical interpretation of the metric \eqref{Boundary-metric-fixed} employed in Sec.~\ref{sec:fluid-gravity}. To this end, let us consider the induced metric on a constant-$r$ hypersurface $\mathcal{B}$,
	\begin{equation}\label{eq:boundary-metric}
		\dif s^2|_{\mathcal{B}}= -r^2\,  (\mathcal{A}^{\text{\tiny L}})^2+  (\mathcal{A}^{\text{\tiny M}})^2   \, .
	\end{equation}
	Its causal character depends explicitly on the radial position. At the Rindler horizon, $r=0$, the hypersurface $\mathcal{B}$ becomes null, and consequently the induced metric \eqref{eq:boundary-metric} is degenerate. Likewise, in the asymptotic region $r\rightarrow+\infty$, where the spacetime is asymptotically locally flat, the conformal class of the induced metric also becomes degenerate, as expected for null infinity.
	
	To isolate the intrinsic conformal geometry of the constant-$r$ slices, we introduce the Weyl-rescaled metric
	\begin{equation}\label{Uni-Mod-Metric}
		\dif \sigma^2 = W^{-1} \dif s^2|_{\mathcal{B}} \, ,
	\end{equation}
	with conformal factor
	\begin{equation}
		W:=\frac{1}{4} \left(\mathcal{P}^2 - r^2 \mathcal{L}^2\right) \, .
	\end{equation}

	This rescaling removes the overall conformal factor of the induced metric while preserving its causal structure and conformal properties. Consequently, geometrical quantities that depend only on the conformal class remain unchanged.
	
	With this choice of Weyl factor, the metric $\gamma_{ab}$ introduced in \eqref{Boundary-metric-fixed}, which plays the role of the background geometry in the fluid/gravity correspondence, is naturally identified with the conformal metric evaluated on the unit-radius hypersurface. Explicitly,
	\begin{equation}\label{gamma-sigma}
		\gamma_{ab}= \sigma_{ab}|_{r=1}\, .
	\end{equation}
	This identification provides a direct geometrical interpretation of $\gamma_{ab}$: it is the representative of the conformal class of the induced metric on constant-$r$ hypersurfaces, normalized by the Weyl factor $W$, and serves as the effective boundary geometry on which the dual fluid variables are defined.
	
	The existence of a well-defined Weyl scaling $W\geq0$ at $r=1$ requires $|\mathcal{P}|\geq|\mathcal{L}|$. This condition has a direct consequence for the allowed range of the energy. In particular, upon using \eqref{QLE}, it implies that the energy is not merely positive but is bounded from below according to
	\begin{equation}
		\mathscr{E} \geq \mathscr{E}_{\text{\tiny min.}} \, ,
	\end{equation}
	where
	\begin{equation}
		\mathscr{E}_{\text{\tiny min.}} :=
		\frac{1}{4}
		\int_{\Sigma}\mathcal{L}^2 \, \dif x \, .
	\end{equation}
	Thus, the requirement of a regular and positive Weyl scaling restricts the physical phase space by excluding configurations whose energy falls below $\mathscr{E}_{\text{\tiny min.}}$.

	Consequently, because $\mathcal{H}_\pm$ are non-negative, the energy is necessarily larger than the absolute value of the momentum,
	\begin{equation}\label{E-M-inequality}
		\mathscr{E} \geq|\mathscr{J}| \, .
	\end{equation}
	Therefore, the requirement of a positive and well-defined Weyl scaling at $r=1$ restricts the physical phase space to configurations satisfying a strict energy--momentum inequality.
	
	The resulting inequality \eqref{E-M-inequality}
	has the form of a dominant-energy bound: the conserved energy strictly dominates the magnitude of the conserved momentum. In the present setting, however, this inequality arises directly from the geometric requirement $W\geq0$ at $r=1$, rather than being imposed as an independent energy condition.

	%%%%%%%%%%%%%%%%%%%%%%%%%%%%%%%%%%%%%%%%%%%%%%%%%%%%%%%%%%%%%%%%%%%%%%%%%%%%
	\section{More on symplectic form}\label{sec:symplectic}
	%%%%%%%%%%%%%%%%%%%%%%%%%%%%%%%%%%%%%%%%%%%%%%%%%%%%%%%%%%%%%%%%%%%%%%%%%%%%
	
	The renormalized symplectic potential and the corresponding symplectic form admit a representation solely in terms of the boundary fields $X$ and $T$,
	\begin{equation}
		\boldsymbol{\Theta}_{\text{\tiny ren.}}
		=  \int_{\partial \mathcal{B}} \delta T \dif X -\frac{1}{2}\, \delta \! \int_{\partial \mathcal{B}}  T \dif X \, ,
	\end{equation}
	and
	\begin{equation}\label{on-shell-S-F}
		\boldsymbol{\Omega}_{\text{\tiny ren.}}
		= - \int_{\partial \mathcal{B}} \delta T \curlywedge \dif \delta X \, .
	\end{equation}
	The second term in the symplectic potential is an exact variation and therefore does not contribute to the symplectic form. Equation~\eqref{on-shell-S-F} consequently provides a direct expression for the renormalized symplectic structure in terms of the boundary data.
	
	The renormalized symplectic form vanishes provided that, for arbitrary variations $\delta_1$ and $\delta_2$, the following condition is satisfied:
	\begin{equation}\label{C-dX-dT}
		\delta_1 \dif T  \wedge \delta_2 \dif  X- \delta_2 \dif T  \wedge \delta_1 \dif  X =0 \, .
	\end{equation}
	This condition, which must hold for every admissible pair $\delta_1,\delta_2$, ensures that the corresponding contribution to the symplectic form vanishes identically on the space of allowed variations.
	
	Using the relations between $X$, $T$, and the boundary currents, the symplectic form can equivalently be written as
	\begin{equation}\label{on-shell-S-F-recast}
		\boldsymbol{\Omega}_{\text{\tiny ren.}}
		= \boldsymbol{\Omega}_{\Sigma(t_f)}-\boldsymbol{\Omega}_{\Sigma(t_i)}+ \boldsymbol{\Omega}_{\text{\tiny end}} \, ,
	\end{equation}
	where
	\begin{subequations}\label{on-shell-S-F-recast-parts}
		\begin{align}
			\boldsymbol{\Omega}_{\Sigma(t)}
			&=
			-\frac{1}{2}
			\int_{\Sigma}
			\delta T\curlywedge\delta\mathcal P\,\dif x,
			\label{on-shell-S-t-cut}
			\\
			\boldsymbol{\Omega}_{\mathrm{end}}
			&=
			2
			\left[
			\int_{t_i}^{t_f}
			\delta T\curlywedge\delta\nu\,\dif t
			\right]_{x=-\infty}^{x=+\infty}.
			\label{eq:endpoint-symplectic-term}
		\end{align}
	\end{subequations}
	This form makes explicit the two types of contributions that arise when the boundary integral is decomposed according to the spacetime directions. The first and second terms are associated with the initial and final constant-time cuts, whereas the third/last is supported on the spatial ends of the boundary.

	On an arbitrary constant-time cut $\Sigma(t)$, the symplectic form induced by \eqref{on-shell-S-F} takes the canonical form \eqref{on-shell-S-t-cut}. Hence, the boundary symplectic structure on a constant-time slice is governed by the pair $(T,\mathcal{P})$. The remaining contribution in \eqref{on-shell-S-F-recast} is a boundary term associated with the spatial ends and is therefore controlled by the boundary conditions imposed there. In particular, whenever the spatial-end contribution vanishes under the adopted falloff conditions, the symplectic structure is completely characterized by the constant-time expression above.
	
	Let us now consider the endpoint contribution $\boldsymbol{\Omega}_{\text{\tiny end}}$. Using the asymptotic behavior of the fields derived in section~\ref{sec:killing-symmetries}, this contribution reduces to
	\begin{equation}
		\boldsymbol{\Omega}_{\text{\tiny end}} = -\frac{1}{2} \,\delta X_0 \curlywedge \delta \mathrm{L}_0    \, ,
	\end{equation}
	where
	\begin{equation}
		X_0:= X(t_f,+\infty)-X(t_i,+\infty) \, .
	\end{equation}
	Here, $X_0$ is required to be a finite parameter specifying the boundary condition. Consequently, its variation is not an independent fluctuation within the chosen phase space. The endpoint contribution is therefore entirely determined by the zero-mode data associated with the asymptotic fields. In particular, no additional information about the detailed bulk profile of the fields enters $\boldsymbol{\Omega}_{\text{\tiny end}}$ once the asymptotic conditions are imposed. This also makes explicit that the endpoint sector captures only the residual global data associated with the boundaries rather than independent local bulk degrees of freedom.
	
	The decomposition \eqref{on-shell-S-F-recast} together with \eqref{on-shell-S-F-recast-parts} is not unique. In particular, one may redistribute a boundary contribution between the cut and endpoint pieces according to
	\begin{subequations}
		\begin{align}
			\tilde{\boldsymbol{\Omega}}_{\Sigma(t)}
			=& \ \boldsymbol{\Omega}_{\Sigma(t)}+\frac{1}{2} \,\delta T(t,+\infty) \curlywedge \delta \mathrm{P}_0 \, ,\\
			\tilde{\boldsymbol{\Omega}}_{\text{\tiny end}} =  & \ \boldsymbol{\Omega}_{\text{\tiny end}}-\frac{1}{2} \,\delta T_0 \curlywedge \delta \mathrm{P}_0 \, ,
		\end{align}
	\end{subequations}
	where
	\begin{equation}
		T_0:= T(t_f,+\infty)-T(t_i,+\infty) \, .
	\end{equation}
	This redefinition leaves the total renormalized symplectic form $\boldsymbol{\Omega}_{\text{\tiny ren.}}$ unchanged and therefore amounts only to a different choice of symplectic decomposition.
	
	With this choice, the cut and endpoint contributions take the form
	\begin{subequations}\label{re-def-S-P-F}
		\begin{align}
			\tilde{\boldsymbol{\Omega}}_{\Sigma(t)}
			=& \ -\frac{1}{2}
			\int_{\Sigma}
			\delta \tilde{T}\curlywedge\delta\mathcal P\,\dif x \, , \label{on-shell-S-t-cut-tilde}\\
			\tilde{\boldsymbol{\Omega}}_{\text{\tiny end}} =  & \ -\frac{1}{2} \left( \delta T_0 \curlywedge \delta \mathrm{P}_0 + \delta X_0 \curlywedge \delta \mathrm{L}_0 \right) \, ,
		\end{align}
	\end{subequations}
	where
	\begin{equation}
		\tilde{T}(t,x)= T(t,x)- T(t,+\infty) \, ,
	\end{equation}
	denotes the shifted field. By construction, this field obeys
	\begin{equation}
		\tilde{T}(t,+\infty)=0 \, , \qquad \tilde{T}(t,-\infty)=-\frac{1}{2} \, \mathrm{L}_0 \, .
	\end{equation}
	The shift is useful because it fixes the value of $\tilde{T}$ at one endpoint and isolates the corresponding zero-mode contribution in the endpoint symplectic form. In particular, the cut symplectic structure now takes the standard canonical form for the pair $(\tilde{T},\mathcal P)$, with the normalization determined by the overall coefficient in \eqref{on-shell-S-t-cut-tilde}. Inverting this symplectic form gives the equal-time Poisson bracket
	\begin{equation}\label{tilde-T-P-algebra}
		\{ \tilde{T}(t,x), \mathcal{P}(t,y) \} =2\, \delta(x-y)\, ,
	\end{equation}
	with the remaining elementary brackets vanishing. Taking an $x$-derivative of \eqref{tilde-T-P-algebra} immediately yields the corresponding bracket involving $\mathcal{L}$ and $\mathcal P$, and hence reproduces the charge algebra \eqref{charge-algebra}. Thus, the charge algebra is encoded directly in the inverse of the cut symplectic form. The agreement between this algebra and the one obtained independently from the canonical charge analysis provides a nontrivial consistency check of the symplectic decomposition and, in particular, confirms that the endpoint contribution has been separated consistently from the local canonical degrees of freedom.
	
	We emphasize that the decomposition \eqref{on-shell-S-F-recast} is not uniquely determined by the definitions \eqref{re-def-S-P-F}. The resulting freedom in the decomposition and its implications deserve further investigation.

	%%%%%%%%%%%%%%%%%%%%%%%%%%%%%%%%%%%%%%%%%%%%%%
	\section{Schr\"odinger operators and transport matrix}\label{app:schrodinger-transport-matrix}
	%%%%%%%%%%%%%%%%%%%%%%%%%%%%%%%%%%%%%%%%%%%%%%
	
	For a spatial coordinate taking values on the non-compact line,
	$x\in\mathbb{R}$, there is in general no preferred closed contour along which
	one can define a monodromy transformation. The appropriate global data are
	therefore obtained by comparing the asymptotic behavior of solutions at the
	two ends of the line. In this setting, it is useful to distinguish between
	the terminology ``Hill equation'' and the more general Schr\"odinger problem:
	strictly speaking, a Hill equation has a periodic coefficient, whereas a
	Schr\"odinger equation with a non-periodic potential on $\mathbb{R}$ is more
	appropriately viewed as a one-dimensional Schr\"odinger equation, or as a
	Hill-type equation when it has the same second-order structure.
	
	Let us assume charge densities $\mathcal{J}_\pm(t,x)$ are given and they belong to the Schwartz space on $\Sigma$. Consider the spectral problem
	\begin{equation}
		\mathrm{S}_{\pm}\psi_\pm= \lambda_{\pm}^2\psi_\pm\, ,
		\label{GHE-new}
	\end{equation}
	where the differential operator is
	\begin{equation}
		\label{Sch-operator-new}
		\mathrm{S}_{\pm}
		=
		-\partial_x^2
		+
		\frac{12\pi}{c_\pm}\,
		\mathcal{J}_{\pm}(t,x)\, ,
	\end{equation}
	with $\lambda_\pm\in\mathbb{C}$ denoting the spectral parameter and
	$c_\pm\in\mathbb{R}$.
	
	Equation \eqref{GHE-new} is a second-order linear differential equation and
	therefore has a two-dimensional solution space for fixed values of
	$\lambda_\pm$. Choose two linearly independent solutions,
	$\psi_\pm^{\text{\tiny (1)}}$ and $\psi_\pm^{\text{\tiny (2)}}$. They can be
	assembled into the fundamental matrix
	\begin{equation}
		\Psi_\pm= 
		\begin{pmatrix}
			\psi_\pm^{\text{\tiny (1)}} & \psi_\pm^{\text{\tiny (2)}} \\[2mm]
			\partial_x \psi_\pm^{\text{\tiny (1)}} & \partial_x \psi_\pm^{\text{\tiny (2)}}
		\end{pmatrix}\,,
	\end{equation}
	where the dependence on $(t,x;\lambda_\pm)$ is understood. The choice of
	fundamental pair is not unique: any nonsingular constant change of basis
	produces another fundamental matrix. Consequently, the matrix relating two
	fundamental bases depends on this choice, while its determinant is fixed by
	the Wronskian normalization.
	
	The Wronskian of the two solutions is
	\begin{equation}
		W_{\pm}(t;\lambda_{\pm})
		=
		\psi_{\pm}^{\text{\tiny (1)}}
		\partial_x
		\psi_{\pm}^{\text{\tiny (2)}}
		-
		\psi_{\pm}^{\text{\tiny (2)}}
		\partial_x
		\psi_{\pm}^{\text{\tiny (1)}},
	\end{equation}
	and is independent of $x$. This follows from the absence of a first-derivative
	term in the differential equation. Equivalently, the coefficient matrix in
	the associated first-order system has vanishing trace. The Wronskian may,
	however, depend on the spectral parameter. A convenient
	normalization of the fundamental solutions can be chosen such that the
	Wronskian takes a prescribed nonzero constant value in $\lambda_\pm$.
	
	To make the propagation problem explicit, the second-order equation can be
	rewritten as a first-order matrix system,
	\begin{equation}
		\partial_x\Psi_\pm=\mathcal{U}_\pm \Psi_\pm\,,
	\end{equation}
	where
	\begin{equation}\label{Cal-U}
		\mathcal{U}_\pm= 
		\begin{pmatrix}
			0 & 1 \\[2mm]
			\frac{12\pi}{c_\pm} \, \mathcal{J}_\pm -\lambda_\pm^2  & 0
		\end{pmatrix}\,.
	\end{equation}
	The matrix $\mathcal{U}_\pm$ is traceless. It follows from the first-order
	system that the determinant of a fundamental matrix is independent of $x$,
	which is the matrix formulation of the constancy of the Wronskian.
	
	The evolution between two finite points $x_1$ and $x_2$ is represented by the
	path-ordered exponential
	\begin{equation}\label{T-x1-x2}
		\Psi_\pm(t,x_2;\lambda_\pm)= \mathrm{U}_\pm(t,x_2,x_1;\lambda_\pm) \Psi_\pm(t,x_1;\lambda_\pm)\, ,
	\end{equation}
	with
	\begin{equation}\label{U-x2-x1}
		\hspace{-0.1 cm}\mathrm{U}_\pm(t,x_2,x_1;\lambda_\pm)= \mathscr{P} \exp\left[\int_{x_1}^{x_2} \mathcal{U}_\pm(t,y;\lambda_\pm)\, \dif y\right] . 
	\end{equation}
	Since $\mathcal{U}_\pm$ is traceless, the corresponding evolution matrix has
	unit determinant whenever it is defined. Moreover, if
	$\mathcal{J}_\pm$ and $c_\pm$ are real and
	$\lambda_\pm^2\in\mathbb{R}$, then $\mathcal{U}_\pm$ is a real traceless
	matrix and the evolution matrix belongs to $SL(2,\mathbb{R})$. 
	
	For a scattering problem, one is interested in the relation between
	solutions adapted to the two asymptotic regions. Since the potential $\mathcal{J}_\pm$ has
	sufficiently well-behaved limits at spatial infinity, one can introduce
	asymptotic bases at $x\rightarrow-\infty$ and $x\rightarrow+\infty$ and
	compare them through the evolution along the line.
	
	Since $\mathcal{J}_\pm$ decays sufficiently rapidly as $|x|\to +\infty$ such that the spectral problem \eqref{GHE-new} approaches the free equations at both ends of the spatial line. In this asymptotic region, the terms involving $\mathcal{J}_\pm$ become negligible, and one consequently has
	\begin{equation}
		\partial_x^2 \psi_\pm + \lambda_\pm^2 \psi_\pm \approx 0 \, , \quad \text{when} \quad |x| \to + \infty \, .
	\end{equation}
	Thus, for real $\lambda_\pm$, the asymptotic solutions are superpositions of right- and left-moving plane waves. More precisely, consider a solution of \eqref{GHE-new} whose asymptotic behavior at the two ends of the non-compact spatial boundary is specified by
	\begin{subequations}
		\begin{align}
			& \psi_\pm \approx \psi^{\text{\tiny (L)}}_\pm \, , \quad \text{when} \quad x \to - \infty \, , \\
			& \psi_\pm \approx \psi^{\text{\tiny (R)}}_\pm \, , \quad \text{when} \quad x \to + \infty \, ,
		\end{align}
	\end{subequations}
	where
	\begin{equation}
		\hspace{-0.2 cm}  \psi^{\text{\tiny (L/R)}}_\pm=\mathscr{A}_{\pm}^{\text{\tiny (L/R)}}(t;\lambda_\pm)\, e^{i\lambda_\pm x}+ \mathscr{B}_{\pm}^{\text{\tiny (L/R)}}(t;\lambda_\pm)\, e^{-i\lambda_\pm x} ,
	\end{equation}
	The labels $(\mathrm{L})$ and $(\mathrm{R})$ refer to the two asymptotic ends of the spatial line rather than to independent solutions of the spectral problem. 
	The coefficients at one end are therefore not, in general, independent of those at the other end: they are related by the global propagation of the solution through the region where $\mathcal{J}_\pm$ is non-vanishing.
	
	The non-compact counterpart of the monodromy matrix is the transfer matrix, which relates the fundamental solutions at the two ends of the spatial domain. In particular, \eqref{T-x1-x2} implies
	\begin{equation}\label{Psi-evolution-new}
		\Psi^{\text{\tiny (R)}}_\pm= \Upsilon_\pm \, \Psi^{\text{\tiny (L)}}_\pm\, ,
	\end{equation}
	where $\Upsilon_\pm(t,x;\lambda_\pm)$ denotes the transfer matrix associated with the corresponding differential equation. Since the transfer matrix is obtained by evolving a fundamental system of solutions, its determinant is equals unity. More generally, the transfer matrix has unit determinant whenever the underlying first-order system is traceless. This property is independent of the normalization of the fundamental solutions, provided the same normalization convention is used at both endpoints. For a real potential and real $\lambda_\pm^2$, the underlying differential equation admits a real fundamental system, and the transfer matrix belongs to $SL(2,\mathbb{R})$.

	We introduce the two-component vector
	\begin{equation}
		\chi_\pm:= \begin{pmatrix}
			\psi_\pm  \\[2mm]
			\partial_x \psi_\pm
		\end{pmatrix}\,.
	\end{equation}
	Differentiating this vector with respect to $x$ and using the corresponding second-order equation for $\psi_\pm$, we recast the spatial equation as a first-order linear system governed by $\mathcal{U}_\pm$. Its fundamental solution matrix propagates the vector $\chi_\pm$ between arbitrary spatial points,
	\begin{equation}\label{chi-x2-x1}
		\chi_\pm(t,x_2;\lambda_\pm)= \mathrm{U}_\pm(t,x_2,x_1;\lambda_\pm)\,  \chi_\pm(t,x_1;\lambda_\pm)\, ,
	\end{equation}
	where $\mathcal{U}_\pm$ and $\mathrm{U}_\pm$ are defined in \eqref{Cal-U} and \eqref{U-x2-x1}, respectively. Thus, $\mathrm{U}_\pm$ encodes the spatial evolution of the solution and provides the natural framework for relating its asymptotic behavior at opposite ends of the spatial domain.
	
	To formulate this relation in terms of asymptotic data, we introduce the coefficient vectors
	\begin{equation}
		\mathscr{C}_\pm^{\text{\tiny (L/R)}} := \begin{pmatrix}
			\mathscr{A}_\pm^{\text{\tiny (L/R)}}  \\[2mm]
			\mathscr{B}_\pm^{\text{\tiny (L/R)}}
		\end{pmatrix}\,.
	\end{equation}
	For $\lambda_\pm\neq 0$, the asymptotic solution vectors can be expressed in terms of these coefficients using the matrix
	\begin{equation}
		\mathrm{M}_{\pm}:=  \begin{pmatrix}
			e^{i \lambda_\pm x} & e^{-i \lambda_\pm x} \\[2mm]
			i \lambda_\pm e^{i \lambda_\pm x} & -i \lambda_\pm e^{-i \lambda_\pm x}
		\end{pmatrix}\,,
	\end{equation}
	so that
	\begin{equation}
		\chi^{\text{\tiny (L/R)}} = \mathrm{M}_{\pm} \mathscr{C}_\pm^{\text{\tiny (L/R)}}\,.
	\end{equation}
	The columns of $\mathrm{M}_\pm$ represent the two independent plane-wave solutions and their spatial derivatives, while $\mathscr{A}_\pm^{\text{\tiny (L/R)}}$ and $\mathscr{B}_\pm^{\text{\tiny (L/R)}}$ specify the amplitudes of these modes in the corresponding asymptotic region. The restriction $\lambda_\pm\neq 0$ ensures that the two columns are linearly independent, so that the coefficient vectors are uniquely defined by the asymptotic solution vectors.
	
	We define the connection between the left- and right-asymptotic coefficients by
	\begin{equation}
		\mathscr{C}_\pm^{\text{\tiny (R)}}= \Xi_\pm \mathscr{C}_\pm^{\text{\tiny (L)}} \, ,
	\end{equation}
	where $\Xi_\pm$ is independent of $x$. This matrix therefore characterizes the transformation of the asymptotic mode coefficients under spatial evolution. The corresponding solution vectors are related by
	\begin{equation}
		\chi^{\text{\tiny (R)}} = \Upsilon_\pm \chi_\pm^{\text{\tiny (L)}}\, ,
	\end{equation}
	where
	\begin{equation}
		\Upsilon_\pm =\mathrm{M}_{\pm} \Xi_\pm \mathrm{M}_{\pm}^{-1} \,.
	\end{equation}
	Hence, $\Xi_\pm$ represents the connection in the plane-wave coefficient basis, whereas $\Upsilon_\pm$ represents the same transformation in the basis of the two-component solution vectors. The two matrices are related by a change of basis and consequently have the same determinant.  \footnote{In the asymptotic regime $|x_1|,|x_2|\gg 1$, the matrix $\Xi_\pm$ can be reconstructed from the known evolution operator $\mathrm U_\pm$ by conjugating it with the corresponding matrices $\mathrm M_\pm$ evaluated at the two endpoints:
		\begin{equation*}
			\Xi_\pm(t;\lambda_\pm)
			=
			\mathrm M_\pm(x_2)^{-1}
			\mathrm U_\pm(t,x_2,x_1;\lambda_\pm)\,
			\mathrm M_\pm(x_1) \, .
		\end{equation*}
		This equation expresses the evolution in the asymptotic frame in terms of the evolution operator in the original frame.}
	
	The transport matrix $\Upsilon_\pm$ belongs to $SL(2,\mathbb{R})$, we accordingly impose $\det\Xi_\pm=1$. These conditions constrain the admissible connection matrices and, in the oscillatory regime, lead to the standard scattering parametrization.
	
	For real, nonzero $\lambda_\pm\in\mathbb{R}\setminus\{0\}$, the reality condition on $\Upsilon_\pm$ translates into the following form of $\Xi_\pm$:
	\begin{equation}
		\Xi_\pm =\begin{pmatrix}
			\bar{\mathscr{T}}^{-1} & -\bar{\mathscr{T}}^{-1} \bar{\mathscr{R}} \\[2mm]
			-\mathscr{T}^{-1} \mathscr{R} & \mathscr{T}^{-1}
		\end{pmatrix} ,
	\end{equation}
	where $\mathscr{R}(t;\lambda_\pm)$ and $\mathscr{T}(t;\lambda_\pm)$ denote the reflection and transmission coefficients, respectively. The unit-determinant condition then gives
	\begin{equation}
		|\mathscr{R}|^2+|\mathscr{T}|^2=1\,.
	\end{equation}
	This is the familiar flux-conservation relation for one-dimensional scattering with real wave number, provided the underlying spatial equation has the usual real, self-adjoint scattering structure and the coefficients are normalized consistently. In this regime, $\Xi_\pm(t;\lambda_\pm)$ is thus expressed in terms of the reflection and transmission data, which encode the mixing of the two asymptotic plane-wave modes.  \footnote{The asymptotic expansion of the logarithmic transmission amplitude provides a generating function for the conserved quantities of the chiral KdV hierarchies. In particular, its first two nonvanishing coefficients encode the integrated chiral currents and their quadratic energies, respectively. The normalization in which the logarithm vanishes as the spectral parameter tends to infinity fixes the relevant branch and removes an additive ambiguity. For smooth Schwartz-class solutions, the isospectral evolution preserves the transmission amplitude, and hence each coefficient in its large-spectral-parameter expansion is time independent. The corresponding chiral charges and energies are therefore conserved under the KdV flow.}
	
	The situation differs when the spectral parameter is purely imaginary, $\lambda_\pm=i\kappa_\pm$, with $\kappa_\pm\in\mathbb{R}\setminus\{0\}$. In this case, $\mathrm{M}_\pm$ is real, and the reality of $\Upsilon_\pm$ implies that $\Xi_\pm$ is real as well. Together with the unit-determinant condition, this gives $\Xi_\pm\in SL(2,\mathbb{R})$. For $\kappa_\pm>0$, the two asymptotic modes behave exponentially. Consequently, its left-asymptotic expansion cannot contain the mode that grows as $x\to-\infty$, and its right-asymptotic expansion cannot contain the mode that grows as $x\to+\infty$. With the coefficient conventions adopted above, this requires
	\begin{equation}
		\mathscr{A}_{\pm}^{\text{\tiny (L)}}=0 \, , \qquad \mathscr{B}_{\pm}^{\text{\tiny (R)}}=0 \, ,
	\end{equation}
	which imposes a nontrivial constraint on the connection matrix $\Xi_\pm$. In particular, a bound state corresponds to a connection between the left- and right-asymptotic coefficient vectors for which the decaying solution on one side evolves into the decaying solution on the other. The resulting connection matrix is therefore not diagonal in general.
	
	The dynamics may be formulated in terms of a pair of operators, $(\mathrm{D}_{\pm},\mathrm{S}_{\pm})$, satisfying the Lax equation \cite{lax1968integrals,ablowitz1981solitons,dickey2003soliton}
	\begin{equation}\label{Lax-eq-new}
		\partial_t \mathrm{S}_{\pm}
		=
		\pm [\mathrm{D}_{\pm},\mathrm{S}_{\pm}]\, .
	\end{equation}
	This commutator evolution is the defining feature of an isospectral flow: under suitable regularity and domain assumptions, the evolution of $\mathrm{S}_{\pm}$ is generated by a similarity transformation and preserves its spectrum. To make this property explicit, we differentiate \eqref{GHE-new} with respect to time and use \eqref{Lax-eq-new}. Assuming that $\lambda_{\pm}$ are independent of time, we obtain
	\begin{equation}
		\mathrm{S}_{\pm}
		\left(
		\partial_t\psi_{\pm}
		\mp
		\mathrm{D}_{\pm}\psi_{\pm}
		\right)
		=
		\lambda_{\pm}^{2}
		\left(
		\partial_t\psi_{\pm}
		\mp
		\mathrm{D}_{\pm}\psi_{\pm}
		\right)
		\, .
	\end{equation}
	Thus, whenever $ \psi_{\pm} $ satisfies the spectral equation \eqref{GHE-new}, the combination $\partial_t\psi_{\pm}\mp\mathrm{D}_{\pm}\psi_{\pm}$ also satisfies the same spectral equation with the same eigenvalue $\lambda_{\pm}^{2}$. This compatibility between the spectral problem and the time evolution establishes the isospectral character of the Lax flow, provided that the relevant operator domains and boundary conditions are preserved during the evolution. In particular, the spectrum of $\mathrm{S}_{\pm}$ remains invariant under the flow, a property that underlies the integrable structure of the dynamics \cite{faddeev1987hamiltonian,dickey2003soliton}.

	%%%%%%%%%%%%%%%%%%%%%%%%%%%%%%%%%%%%%%%%%%%%%%%%%%%%%%%%%%%%%%%%%%%%%%%%%%%%
	%\section{Different boundary conditions}
	%%%%%%%%%%%%%%%%%%%%%%%%%%%%%%%%%%%%%%%%%%%%%%%%%%%%%%%%%%%%%%%%%%%%%%%%%%%%
	%In this section we present some boundary conditions.
	
	%%%%%%%%%%%%%%%%%%%%%%%%%%%%%%%%%%%%%%%%%%%%%%%%%%%%%%%%%%%%
	\section{Heisenberg-like boundary conditions}\label{Heisenberg-BC}
	%%%%%%%%%%%%%%%%%%%%%%%%%%%%%%%%%%%%%%%%%%%%%%%%%%%%%%%%%%%%
	
	It follows from \eqref{C-E-M-C} and \eqref{conservation-condition} that the simplest choice ensuring conservation of both energy and momentum is to take $\mu_\pm$ to be constant. In this case, the renormalized symplectic form \eqref{renormalized-symplectic-form}simplifies to
	\begin{equation}
		\boldsymbol{\Omega}_{\text{\tiny ren.}}
		= -  \left( \delta \mu \curlywedge \delta \mathrm{P}_0 +\delta \nu \curlywedge  \delta \mathrm{L}_0 \right) (t_f-t_i) \,  ,
	\end{equation}
	which makes explicit that the nontrivial phase-space pairing is entirely encoded in the zero modes $\mathrm P_0$ and $\mathrm L_0$ and their conjugate parameters $\mu$ and $\nu$.
	
	The equations of motion \eqref{eq:EOM-RPS} then imply that the dynamical fields $\mathcal L$ and $\mathcal P$ are independent of time,
	\begin{equation}
		\mathcal{L} = \mathcal{L}(x) \, , \qquad
		\mathcal{P}  =\mathcal{P}(x) \, .
	\end{equation}
	Consequently, the corresponding potentials have at most a linear dependence on time,
	\begin{subequations}
		\begin{align}
			& T(t,x)= T_\circ (x)+2\, \mu\,  t \, , \\
			& X(t,x)= X_\circ (x)+2\, \nu\, t \, ,
		\end{align}
	\end{subequations}
	where $T_\circ$ and $X_\circ$ are time-independent functions. The same structure is inherited by the symmetry parameters, which can be written as
	\begin{subequations}
		\begin{align}
			& \zeta(t,x)= \zeta_\circ (x)+ \check{\zeta}\, t \, , \\
			& \xi(t,x)= \xi_\circ (x)+ \check{\xi}\, t \, ,
		\end{align}
	\end{subequations}
	with $\zeta_\circ$ and $\xi_\circ$ denoting their time-independent parts.
	
	The coefficients $\check{\zeta}$ and $\check{\xi}$ correspond to transformations that do not carry an associated surface charge. They therefore represent proper gauge directions in phase space and may consistently be set to zero. The residual transformations are consequently characterized by
	\begin{subequations}
		\begin{align}
			& \delta_\epsilon \mathcal{L} = 4 \,\partial_x \zeta_\circ  \, , & & 
			\delta_\epsilon \mathcal{P}  = 4 \,\partial_x\xi_\circ  \, , \\
			&\delta_\epsilon \mu =   0 \, , & & 
			\delta_\epsilon \nu = 0 \, .
		\end{align}
	\end{subequations}
	This should be distinguished from a parametric variation in phase space, for which $\delta\mu$ and $\delta\nu$ need not vanish.
	
	To specify the phase space further, we impose the following functional relation between the chemical potentials and the conserved zero modes:
	\begin{equation}\label{cal-F-def}
		\mu = \frac{\partial \mathscr{F}}{\partial \mathrm{P}_0} \, , \qquad  \nu = \frac{\partial \mathscr{F}}{\partial \mathrm{L}_0} \, ,
	\end{equation}
	where $\mathscr{F}=\mathscr{F}(\mathrm{P}_0,\mathrm{L}_0)$ is an otherwise arbitrary differentiable function. This condition reduces the independent phase-space data by expressing the chemical potentials as functions of the conserved charges.
	
	A further restriction on $\mathscr F$ follows if one requires the renormalized symplectic potential $\boldsymbol{\Theta}_{\text{\tiny ren.}}$ to vanish. Indeed, using \eqref{renormalized-symplectic-potential}, this requirement gives
	\begin{equation}\label{Free-energy-eq}
		2\mathscr{F} = \frac{\partial \mathscr{F}}{\partial \mathrm{P}_0}  \, \mathrm{P}_0 +\frac{\partial \mathscr{F}}{\partial \mathrm{L}_0} \,   \mathrm{L}_0 \, .
	\end{equation}
	This is precisely Euler's homogeneity condition of degree two. Hence, $\mathscr F$ must be homogeneous of degree two in $(\mathrm P_0,\mathrm L_0)$. On the patch $\mathrm P_0\neq0$, the general differentiable solution may therefore be parametrized as
	\begin{equation}
		\mathscr{F}(\mathrm P_0,\mathrm L_0) = \mathrm P_0^2\, f\!\left(\mathrm L_0/\mathrm P_0\right) ,
	\end{equation}
	for an arbitrary differentiable function $f$. This representation makes the degree-two homogeneity manifest, since the dependence on the overall scale of the charges is carried entirely by the prefactor $\mathrm P_0^2$, while their relative magnitude is encoded in the ratio $\mathrm L_0/\mathrm P_0$.
	
	Varying $\mathscr{F}$ gives the first law in the form
	\begin{equation}
		\delta \mathscr{F}= \mu \,\delta \mathrm P_0+\nu \,\delta \mathrm L_0 \, .
	\end{equation}
	This relation admits a thermodynamic interpretation upon identifying $\mathscr{F}$ with the free--energy, $\pi \mathrm{P}_0$ with the entropy, and $\mathrm{L}_0$ with the momentum. The corresponding intensive variables are then identified as the temperature $\mu/\pi$ and the potential $\nu$ conjugate to the momentum. Accordingly, $\mathscr{F}$ is interpreted as a thermodynamic potential whose variations with respect to the entropy and momentum are governed by the temperature and its conjugate potential, respectively.

	%%%%%%%%%%%%%%%%%%%%%%%%%%%%%%%%%%%%%%%%%%%%%%%%%%%%%%%%%%%%
	\section{Dirichlet-like boundary conditions}\label{Dirichlet-BC}
	%%%%%%%%%%%%%%%%%%%%%%%%%%%%%%%%%%%%%%%%%%%%%%%%%%%%%%%%%%%%
	
	Motivated by the hydrodynamic description developed in section \ref{sec:fluid-gravity}, we impose Dirichlet boundary conditions by fixing the chemical potentials according to
	\begin{subequations}
		\begin{align}
			& \mu = \frac{1}{4}\left(
			N_\circ \, \mathcal{P} + N^x_\circ \, \mathcal{L}
			\right), \\
			& \nu = \frac{1}{4}\left(
			N_\circ\, \mathcal{L} +N^x_\circ\,  \mathcal{P}
			\right) \, .
		\end{align}
	\end{subequations}
	Here, $N_\circ$ and $N^x_\circ$ are spacetime-independent constants. Equivalently, in terms of the diagonal variables, these boundary conditions take the form
	\begin{equation}\label{D-L-BC}
		\mu_\pm=\frac{1}{2}\, v_\pm \, \mathcal{J}_\pm \, , \quad \text{with} \quad v_\pm= N_\circ \pm N^x_\circ \, .
	\end{equation}
	Up to a constant normalization and/or a constant shift of the time and spatial coordinates, the resulting evolution can thus be viewed as propagation on a Minkowski background.
	
	The conservation condition \eqref{conservation-condition}, together with the above boundary conditions, requires the asymptotic values of the chiral currents to agree at the two ends of the non-compact spatial direction,
	\begin{equation}
		\mathcal{J}_\pm(t,+ \infty)=\mathcal{J}_\pm(t,- \infty) \, .
	\end{equation}
	This condition eliminates the corresponding boundary flux and, consequently, ensures conservation of both the total energy \eqref{QLE} and the total momentum \eqref{QLM}.

	For the Dirichlet-like boundary conditions specified above, the renormalized symplectic form reduces to
	\begin{equation}
		\boldsymbol{\Omega}_{\text{\tiny ren.}} = -  \left( \delta N_\circ \curlywedge \delta \mathscr{E} -\delta N^x_\circ \curlywedge \delta \! \mathscr{J} \right) \left(t_f-t_i \right) \, .
	\end{equation}
	Consequently, the symplectic flux vanishes if the variations of $N_\circ$ and $N^x_\circ$ are integrable on the space of solutions. In particular, suppose that there exists a differentiable function $\mathscr{S}(\mathscr{E},\mathscr{J})$ such that
	\begin{equation}
		N_\circ = \frac{1}{2\pi}\, \frac{\partial \mathscr{S}}{\partial \mathscr{E}} \, , \qquad N^x_\circ = -\frac{1}{2\pi}\, \frac{\partial \mathscr{S}}{\partial \!\mathscr{J}} \, .
	\end{equation}
	If, in addition, the renormalized symplectic potential is required to vanish, the corresponding condition is
	\begin{equation}\label{entropy-DLBDY}
		\mathscr{S}(\mathscr{E},\mathscr{J})=  \frac{\partial \mathscr{S}}{\partial \mathscr{E}} \, \mathscr{E} + \frac{\partial \mathscr{S}}{\partial \!\mathscr{J}} \mathscr{J} \, ,
	\end{equation}
	which is Euler's equation for a function homogeneous of degree one. Thus, $\mathscr{S}$ must be homogeneous of degree one in the energy and momentum. Its general differentiable solution can be expressed as
	\begin{equation}\label{F-function}
		\mathscr{S}(\mathscr{E},\mathscr{J}) = \mathscr{E}\,h\!\left(\frac{\mathscr{J}}{\mathscr{E}}\right)\, ,
	\end{equation}
	with $h$ an arbitrary differentiable function. The representation in \eqref{F-function} is local in the $(\mathscr{E},\mathscr{J})$ plane and should therefore be understood on a patch for which the ratio $\mathscr{J}/\mathscr{E}$ is well defined.
	
	Taking the variation of \eqref{F-function} and using \eqref{E-J-H-p-m}, we obtain
	\begin{equation}
		\delta \mathscr{E}= \vartheta \, \delta \mathscr{S}+\varpi \,  \delta\!\mathscr{J} \, .
	\end{equation}
	This equation has precisely the structure of the first law of thermodynamics, with
	\begin{equation}
		\vartheta :=\frac{1}{2\pi N_\circ} \, , \qquad \varpi :=\frac{N^x_\circ}{N_\circ} \, ,
	\end{equation}
	playing the roles of temperature and chemical potential conjugate variable, respectively. The first-law interpretation is consequently a statement about the thermodynamic form of the variation of the conserved energy: $\mathscr{S}$ is the entropy, $\vartheta$ its conjugate intensive parameter, and $\varpi$ the intensive quantity conjugate to $\mathscr{J}$. This interpretation does not require $\mathscr{S}$ to be identified with a thermodynamic entropy independently of the variational analysis; rather, the identification follows from the integrability and homogeneity conditions imposed above.

	%%%%%%%%%%%%%%%%%%%%%%%%%%%%%%%%%%%%%%%%%%%%%%%
	\subsection{Generic solutions and Lax pair}
	%%%%%%%%%%%%%%%%%%%%%%%%%%%%%%%%%%%%%%%%%%%%%%%
	
	For the chemical potentials specified in \eqref{D-L-BC}, the equations of motion \eqref{EOM-chiral} decouple into two independent chiral evolution equations. Their general local solutions can therefore be expressed as $\mathcal{J}_\pm=\mathcal{J}_\pm(z^\pm)$, or equivalently as $\Phi_\pm=\Phi_\pm(z^\pm)$, where $z^\pm=x\pm v_\pm t$. Thus, the two chiral sectors evolve independently, with each sector depending only on its corresponding characteristic coordinate. The requirement that the symmetry transformations preserve this chiral structure likewise restricts the symmetry parameters to the form $\eta_\pm=\eta_\pm(z^\pm)$.
	
	To construct the associated Lax pair, we choose
	\begin{equation}
		\mathrm{D}_\pm = v_\pm \partial_x \, , 
	\end{equation}
	which satisfies the Lax equation \eqref{Lax-eq-new}. The auxiliary linear evolution equations $\partial_t\psi_{\pm}=\pm\mathrm{D}_{\pm}\psi_{\pm}$ then imply that $\psi_\pm$ depends on spacetime through the corresponding characteristic coordinate, $\psi_\pm=\psi_\pm(z^\pm)$. Consequently, the Schr\"odinger equations \eqref{GHE-new} can be written in the form
	\begin{equation}\label{Sch-eq-D}
		-\partial_\pm^2 \psi_\pm
		+
		\frac{12\pi}{c_\pm}\,
		\mathcal{J}_{\pm}\psi_\pm = \lambda_{\pm}^2\psi_\pm\, ,
	\end{equation}
	where $\partial_\pm$ denotes differentiation with respect to $z^\pm$. In this representation, each chiral sector is governed by a second-order spectral problem whose potential is determined by the corresponding current $\mathcal{J}_\pm$.

	%%%%%%%%%%%%%%%%%%%%%%%%%%%%%%%%%%%%%%%%%%%%%%%
	\subsection{P\"oschl--Teller system}
	%%%%%%%%%%%%%%%%%%%%%%%%%%%%%%%%%%%%%%%%%%%%%%%
	We consider a class of chiral fields characterized by a hyperbolic-tangent profile,
	\begin{equation}
		\Phi_\pm(z^\pm)=C_\pm + \frac{1}{4 } \, \mathrm{J}_0^\pm\, \tanh\!{\left[ \frac{12\, \mathscr{H}_\pm\,  z^\pm}{(\mathrm{J}_0^\pm)^2}\right]} \, ,
	\end{equation}
	where $C_\pm$ are integration constants. Differentiating these fields with respect to their respective chiral coordinates yields the chiral currents
	\begin{equation}\label{J-Sech-ex}
		\mathcal{J}_\pm(z^\pm) =   \frac{6\,\mathscr{H}_\pm}{\mathrm{J}_0^\pm}\, \sech^2\!{\left[ \frac{12\, \mathscr{H}_\pm\,  z^\pm}{(\mathrm{J}_0^\pm)^2}\right]}\, .
	\end{equation}
	For nonvanishing $\mathrm{J}_0^\pm$ and $\mathscr{H}_\pm$, the current profiles are localized, approaching zero as $z^\pm\to\pm\infty$. Note that $\mathrm{J}_0^\pm$ and $\mathscr{H}_\pm$ are independent. Their characteristic width is controlled by the ratio $(\mathrm{J}_0^\pm)^2/\mathscr{H}_\pm$, while their overall amplitude is determined by $\mathscr{H}_\pm/|\mathrm{J}_0^\pm|$. Thus, the currents describe localized pulses with a $\sech^2$ profile, whereas the corresponding chiral fields interpolate smoothly between two asymptotic values.
	
	In particular, the asymptotic behavior of the chiral fields is
	\begin{subequations}
		\begin{align}
			& \Phi_\pm(t,+ \infty)= C_\pm + \frac{1}{4 }  \mathrm{J}_0^\pm \, ,\\
			& \Phi_\pm(t,- \infty)= C_\pm - \frac{1}{4 }  \mathrm{J}_0^\pm \, ,
		\end{align}
	\end{subequations}
	so that the difference between the two limiting values is fixed by the zero mode $\mathrm{J}_0^\pm$, independently of the integration constant $C_\pm$. The latter determines the average of the two asymptotic values. The parameters $C_\pm$ and $\mathrm{J}_0^\pm$ therefore control the offset and the net change of the chiral fields, respectively, while the Hamiltonians $\mathscr{H}_\pm$ set the scale of the localized transition through the profile's width. The constants and zero modes must be specified consistently with the chosen phase-space sector and the definitions of the chiral Hamiltonians.
	
	We now introduce the parametrization
	\begin{equation}
		\mathrm{J}_0^\pm= -\beta_\pm   \mathscr{H}_\pm^{\frac{1}{3}}
	\end{equation}
	where $\beta_\pm>0$ are constant parameters. To recast the Schr\"odinger equation \eqref{Sch-eq-D} into a standard form, we define the rescaled coordinates
	\begin{equation}
		Z_\pm = \frac{12}{\beta_\pm^2   }\, \mathscr{H}_\pm^{\frac{1}{3}} z^\pm \, ,
	\end{equation}
	and choose the constants
	\begin{equation}
		c_\pm = \frac{\pi \beta_\pm^3}{2 l_\pm (l_\pm +1)}\, , 
	\end{equation}
	where $l_\pm$ are parameters characterizing the strength of the resulting potential. With these choices, equation \eqref{Sch-eq-D} takes the form
	\begin{equation}
		-  \frac{\dif{}^2 \psi_\pm}{\dif Z_\pm^2}
		-
		l_\pm (l_\pm +1) \sech^2\!{\left( Z_\pm\right)}\, \psi_\pm = \varepsilon_\pm^2\,\psi_\pm\, ,
	\end{equation}
	which is the stationary Schr\"odinger equation for the attractive P\"oschl--Teller potential. The corresponding spectral parameter is
	\begin{equation}
		\varepsilon_\pm = \frac{\beta_\pm^2   \,\lambda_{\pm}}{12 \mathscr{H}_\pm^{\frac{1}{3}}} \, .
	\end{equation}
	
	The rescaling isolates the dependence on $\mathscr{H}_\pm$ in the dimensionless spectral parameter $\varepsilon_\pm$, while the shape of the potential is governed by $l_\pm$. In particular, the potential is reflectionless when $l_\pm$ is a non-negative integer, yielding a useful connection between the original equation and the exactly solvable P\"oschl--Teller system. For generic values of $l_\pm$, the equation remains a P\"oschl--Teller problem, although the reflectionless property does not generally hold.

	%%%%%%%%%%%%%%%%%%%%%%%%%%%%%%%%%%%%%%%%%%%%%%%%%%%%%%%%%%%%%%%%%%%%%%%%%%%%
	\section{Quantization}\label{sec:quantization}
	%%%%%%%%%%%%%%%%%%%%%%%%%%%%%%%%%%%%%%%%%%%%%%%%%%%%%%%%%%%%%%%%%%%%%%%%%%%%
	
	We now discuss the quantization of the boundary phase space associated with the
	fields $\mathcal{L}(t,x)$ and $\mathcal{P}(t,x)$ on the noncompact spatial
	slice $\Sigma=\mathbb{R}$.
	The noncompactness of $\Sigma$ is important for the quantization, since the
	Fourier decomposition involves a continuous momentum variable rather than a
	discrete mode number. Moreover, the treatment of global zero modes requires
	some care and should be distinguished from the continuum of local oscillator modes. Classically, they are obtained by integrating the currents over the entire spatial slice. Their quantum definition will require a separate prescription.
	
	In the representation constructed below, the classical phase-space variables $\mathcal{L}$ and $\mathcal{P}$ are promoted to operator-valued distributions, denoted by $\hat{\mathcal{L}}$ and $\hat{\mathcal{P}}$, respectively. Following Dirac's canonical quantization prescription, the classical Poisson brackets are replaced by equal-time commutators according to $\{\,\cdot\,,\,\cdot\,\}\rightarrow [\,\cdot\,,\,\cdot\,]/i$. The resulting relations are understood in the distributional sense, with the spatial arguments $x$ and $y$ smeared independently against smooth, compactly supported test functions.
	
	Applying this prescription to the classical bracket between $\mathcal{L}$ and $\mathcal{P}$ gives the equal-time quantum commutator
	\begin{equation}\label{CECR}
		[
		\hat{\mathcal{L}}(t,x),
		\hat{\mathcal{P}}(t,y)]
		=
		4i\,
		\partial_x
		\delta(x-y)\, .
	\end{equation}
	The commutators of $\hat{\mathcal L}$ with itself and of $\hat{\mathcal P}$ with itself vanish.
	
	It is convenient to express the algebra in terms of the chiral combinations $\hat{\mathcal J}_{\pm}=(\hat{\mathcal P}\pm\hat{\mathcal L})/2$. Their equal-time commutators are
	
	\begin{equation}\label{CCR-JJ}
		\begin{aligned}
			\bigl[\hat{\mathcal J}_{\pm}(t,x),\hat{\mathcal J}_{\pm}(t,y)\bigr]
			&=\pm 2i\,\partial_x\delta(x-y)\,,\\
			\bigl[\hat{\mathcal J}_{+}(t,x),\hat{\mathcal J}_{-}(t,y)\bigr]
			&=0\,.
		\end{aligned}
	\end{equation}
	The two chiral current algebras commute with each other and therefore constitute independent current algebras. The term proportional to the derivative of the Dirac delta function represents the central extension of the chiral algebra at the quantum level. The relative sign between the two sectors reflects their opposite chiralities and the corresponding sign difference in their current algebras.
	
	Since $\hat{\mathcal{L}}$, $\hat{\mathcal{P}}$, and $\hat{\mathcal{J}}_{\pm}$ are operator-valued distributions, products of these operators at coincident points are, in general, not well-defined without further specification. Consequently, nonlinear composite operators require an appropriate regularization prescription and, where necessary, a definite choice of operator ordering. The distributional current algebra itself is linear and requires only smearing. An additional prescriptions enters when we form nonlinear composite operators.

	%%%%%%%%%%%%%%%%%%%%%%%%%%%%%%%%%%%%%%%%%%%
	\subsection{Setup on the real line}\label{sec:XVA}
	%%%%%%%%%%%%%%%%%%%%%%%%%%%%%%%%%%%%%%%%%%%
	
	\begin{definition}
		Let $\mathbb{T}(\Sigma)=C_0^\infty(\Sigma)$ denote the real vector space of smooth compactly supported test functions on the non-compact spatial slice $\Sigma$. One may alternatively work with a suitable real Schwartz space when the corresponding falloff conditions are imposed. At fixed time $t$, for $\eta_\pm\in\mathbb{T}(\Sigma)$ define
		\begin{equation}
			\hat{\mathrm J}_\pm(\eta_\pm) := \int_{\Sigma} \dif x \, \eta_\pm(t,x)\,\hat{\mathcal{J}}_\pm(t,x)\, .
		\end{equation}
	\end{definition}
	At this stage, these expressions denote formal smeared current generators. In a regular representation of the Weyl algebra introduced below, they are realized as self-adjoint generators of one parameter unitary groups. In the Fock representation, their commutations relations hold on a common invariant dense domain. 
	
	Smearing the local current algebra with compactly supported test functions removes the distributional singularities and yields the well-defined commutator
	\begin{equation}\label{eq:smeared-CCR}
		\big[\hat{\mathrm J}_\pm(\eta^\pm_1),\hat{\mathrm J}_\pm(\eta^\pm_2)\big]
		= i\,\sigma_\pm(\eta^\pm_1,\eta^\pm_2)\, ,
	\end{equation}
	with
	\begin{equation}
		\sigma_\pm(\eta^\pm_1,\eta^\pm_2) := \pm\, 2\int_{\Sigma} \dif x\,  \eta^\pm_1\,\partial_x\eta^\pm_2\, ,
	\end{equation}
	where the right-hand side of the commutator is proportional to the identity operator and is therefore central. The bilinear form $\sigma_\pm$ is real and antisymmetric. The latter property follows, for compactly supported test functions, by integration by parts, since the boundary term at spatial infinities vanish.
	
	Combining the two chiral sectors, let $\epsilon=(\eta_+,\eta_-)$ and define
	\begin{equation}
		\hat{Q}(\epsilon) := \hat{\mathrm J}_+(\eta_+)+\hat{\mathrm J}_-(\eta_-)\, .
	\end{equation}
	The corresponding central form is
	\begin{equation}\label{eq:sigma}
		\sigma(\epsilon_1,\epsilon_2)
		= 2\int_{\Sigma} \dif x
		\left(\eta_{1}^+\partial_x\eta_{2}^+-\eta_{1}^-\partial_x\eta_{2}^-\right)\, ,
	\end{equation}
	which is real and antisymmetric. Moreover, on $\mathbb{T}(\Sigma)\oplus\mathbb{T}(\Sigma)$ this form is nondegenerate. Indeed, if $\sigma(\epsilon_1,\epsilon_2)=0$ for every $\epsilon_2$, then, by choosing the two chiral components of $\epsilon_2$ independently and integrating by parts, one finds that $\epsilon_1$ must be constant. Since the test functions are compactly supported, this constant must vanish. Hence the pair $\left(\mathbb{T}(\Sigma)\oplus\mathbb{T}(\Sigma),\,\sigma\right)$
	defines an infinite-dimensional symplectic vector space.
	
	The smeared generators consequently satisfy
	\begin{equation}\label{CCRQV}
		[\hat{Q}(\epsilon_1),\hat{Q}(\epsilon_2)]=i\, \sigma(\epsilon_1,\epsilon_2) \, ,
	\end{equation}
	which defines an infinite-dimensional Heisenberg-type canonical commutation relation algebra. The nontrivial central extension is completely determined by the symplectic form $\sigma$. In particular, the two chiral sectors contribute independently, with opposite signs in the central form, while the spatial coordinate remains continuous over the non-compact line $\Sigma$. This formulation therefore provides the natural Weyl-algebraic starting point for the subsequent GNS construction of the quantum theory.
	
	Nonzero constant functions do not belong to $\mathbb T(\Sigma)$. Consequently, the formal integrated currents, corresponding to smearing with the identity, are not among the smeared generators defined here. Their realization as quantum observables requires additional specification, such as a suitable limiting prescription in a chosen representation. This issue does not affect the nondegeneracy of the symplectic form on the chosen space of test functions.
	%%%%%%%%%%%%%%%%%%%%%%%%%%%%%%%%%%%%%%%%%%
	\subsection{The Weyl $C^*$-algebra}
	%%%%%%%%%%%%%%%%%%%%%%%%%%%%%%%%%%%%%%%%%%
	
	Since the charge operators $\hat{Q}(\epsilon)$ are, in general, unbounded,
	the polynomial algebra generated by them is naturally an unbounded
	$*$-algebra. Its formulation therefore involves domain questions and,
	when applicable, questions concerning self-adjoint extensions. A
	convenient way to avoid these representation-dependent issues at the
	algebraic level is to work instead with the exponentiated charges. This
	leads to the Weyl formulation of the canonical commutation relations.
	
	\begin{definition}\label{def:weyl}
		For $\epsilon=(\eta_+,\eta_-)$ in the symplectic space
		$
		\left(\mathbb{T}(\Sigma)\oplus\mathbb{T}(\Sigma),\,\sigma\right),
		$
		introduce formal unitaries $\hat{W}(\epsilon)$ subject only to the
		``Weyl relations''
		\begin{subequations}\label{eq:Weyl-rel}
			\begin{align}
				& \hat{W}(\epsilon_1)\,\hat{W}(\epsilon_2)
				= e^{-\frac i2\sigma(\epsilon_1,\epsilon_2)}\,\hat{W}(\epsilon_1+\epsilon_2)\, , \\
				& \hat{W}(\epsilon)^* = \hat{W}(-\epsilon)\, , \\
				& \hat{W}(0)=\hat{1} \, .
			\end{align}
		\end{subequations}
		In a regular Hilbert-space representation $\pi$, the map $\tau\longmapsto\pi\big(\hat W(\tau\epsilon)\big)$ is strongly continuous for every fixed $\epsilon$. Stone's theorem therefore provides a unique self-adjoint generator $\hat{Q}(\epsilon)$ such that
		\begin{equation}
			\pi \left( \hat{W}(\tau \epsilon) \right):=e^{i\tau\hat{Q}(\epsilon)}, \qquad \tau\in\mathbb R.\, .
		\end{equation}
		The Weyl relations are imposed at the algebraic level. In the Fock representation constructed below, the corresponding generators satisfy the canonical commutation relations specified in
		\eqref{CCRQV} on a common invariant dense domain. The abstract Weyl algebra
		\begin{equation}
			\mathfrak{A}=\mathfrak{A}\left(\mathbb{T}(\Sigma)\oplus\mathbb{T}(\Sigma),\,\sigma\right)\, ,
		\end{equation}
		is the universal unital $C^*$-algebra generated by the symbols
		$\{\hat{W}(\epsilon)\}$ subject to \eqref{eq:Weyl-rel}. The existence of
		this universal completion follows from the general theory of Weyl $C^*$-algebras, commonly associated with Slawny's construction.
	\end{definition}
	
	The Weyl formulation separates the abstract algebraic structure from
	the choice of Hilbert-space representation. In particular, the elements
	$\hat{W}(\epsilon)$ are represented by bounded unitaries in every unital $*$-representation
	of $\mathfrak{A}$, whereas the generators $\hat{Q}(\epsilon)$, when they
	exist, are generally unbounded and their domains depend on the chosen representation. Consequently, states and their GNS representations can be constructed without first specifying a particular realization of the charge operators. Thermal equilibrium additionally requires a specified dynamics, while modular theory applies to the represented von Neumann algebras under the appropriate faithfulness assumptions. As explained in Definition \ref{def:weyl}, the charge operators are recovered through Stone’s theorem in a regular representation. They are therefore regarded as infinitesimal generators of the represented Weyl operators rather than as elements of the $C^*$-algebra itself.
	
	For $\Sigma=\mathbb{R}$, the symplectic space is infinite-dimensional and the Fourier transform carries a continuous momentum label, in contrast to the discrete spectrum associated with periodic boundary conditions on a compact spatial slice. This continuous labeling is a property of the spectral decomposition and does not by itself specify a particular Hilbert-space representation of the Weyl algebra.
	
	The plane waves on $\mathbb{R}$ are not elements of $L^2(\mathbb{R})$ but generalized eigenvectors, understood distributionally. One may therefore work either in a rigged Hilbert-space framework or, equivalently for the present purposes, with normalizable wave packets.
	
	The noncompactness of $\Sigma$ does not alter the abstract Weyl-algebraic construction, but it affects its representations and thermal interpretation. In particular, the existence of a KMS state on the $C^*$-algebra does not imply the existence of a trace-class Gibbs density operator in a given representation. For noncompact $\Sigma$, the latter is generally obstructed by the infinite spatial volume. Thermal equilibrium is therefore most naturally formulated algebraically, with the corresponding GNS representation constructed from the KMS state.
	
	%%%%%%%%%%%%%%%%%%%%%%%%%%%%%%%%%%%%%%%%%%%%%%%%%
	\subsection{States and the GNS construction}
	%%%%%%%%%%%%%%%%%%%%%%%%%%%%%%%%%%%%%%%%%%%%%%%%%
	
	\begin{definition}
		Let $\mathfrak{A}$ be a unital $C^*$-algebra. A \emph{state} on $\mathfrak{A}$ is a linear functional
		$\omega:\mathfrak{A}\to\mathbb{C}$ satisfying
		\begin{enumerate}[label=(\roman*)]
			\item positivity:
			$\omega(\hat{A}^*\hat{A})\geq 0$ for every $\hat{A}\in\mathfrak{A}$;
			\item normalization:
			$\omega(\hat{1})=1$.
		\end{enumerate}
		For a self-adjoint observable $\hat A\in\mathfrak A$, the quantity $\omega(\hat{A})$ is interpreted as the expectation value in the state $\omega$. This definition is purely algebraic and does not require a choice of Hilbert-space representation.
	\end{definition}
	
	We consider \emph{centered quasi-free}, (or \emph{Gaussian}), states on the Weyl algebra, whose expectation values on the Weyl generators are completely determined by a real, symmetric covariance form. In the present notation,
	\begin{equation}\label{quasi-free}
		\omega\left(\hat{W}(\epsilon)\right) = e^{-\frac{1}{2} S(\epsilon,\epsilon)}\, , 
	\end{equation}
	where $S$ is a real, symmetric, non-negative bilinear form on
	$\mathbb{T}(\Sigma)\oplus\mathbb{T}(\Sigma)$. The covariance form must be compatible with the symplectic form $\sigma$ entering the Weyl relations. In particular, it satisfies
	\begin{equation}\label{eq:mu-sigma-ineq}
		S(\epsilon_1,\epsilon_1)
		S(\epsilon_2,\epsilon_2)
		\geq
		\frac{1}{4}
		|\sigma(\epsilon_1,\epsilon_2)|^2 \, ,
	\end{equation}
	for all $\epsilon_1,\epsilon_2$. This inequality expresses the uncertainty relation between the symmetric covariance and the symplectic form.
	
	The quasi-free class is especially useful in the present setting because its correlation functions are determined by the corresponding two-point function through Wick's rule. The centered vacuum constructed below belongs to this class. These states are regular, so their GNS representations admit self-adjoint smeared current generators.
	
	\begin{theorem}[GNS]\label{thm:GNS}
		Let $\mathfrak{A}$ be a unital $C^*$-algebra and let $\omega$ be a state on $\mathfrak{A}$. Then there exists a triple $(\mathcal{H}_\omega,\pi_\omega,\Omega_\omega)$, unique up to unitary equivalence, consisting of
		\begin{itemize}
			\item a Hilbert space $\mathcal{H}_\omega$,
			\item a $*$-representation $\pi_\omega:\mathfrak{A}\to\mathscr{B}(\mathcal{H}_\omega)$,
			\item a unit \emph{cyclic vector} $\Omega_\omega\in\mathcal{H}_\omega$, satisfying
			\begin{equation}
				\overline{\pi_\omega(\mathfrak{A})\Omega_\omega}
				=\mathcal{H}_\omega \, ,
			\end{equation}
		\end{itemize}
		such that
		\begin{equation}
			\omega(\hat{A})
			=
			\langle\Omega_\omega,\pi_\omega(\hat{A})\,\Omega_\omega\rangle
			\qquad\text{for all } \hat{A}\in\mathfrak{A}\, .
		\end{equation}
	\end{theorem}
	
	Theorem~\ref{thm:GNS} is the precise algebraic sense in which
	choosing a state $\omega$ builds a Hilbert space representation. Once a positive normalized functional has been specified, the theorem guarantees the existence and uniqueness, up to unitary equivalence, of its cyclic representation. It does not itself select the state. For the current algebra on the line, the standard vacuum and finite-temperature quasi-free KMS states give inequivalent global GNS representations. Thus, the same abstract observable algebra accommodates vacuum and thermal equilibrium through distinct representations.
	
	Once the GNS representation has been chosen, the represented \(C^*\)-algebra generates a von Neumann algebra by taking its closure in the weak operator topology. By the bicommutant theorem,
	\begin{equation}
		\mathcal M_\omega
		=
		\overline{\pi_\omega(\mathfrak A)}
		=
		\pi_\omega(\mathfrak A)'',
	\end{equation}
	where the overline denotes this closure. The same construction applies to local observables. For a bounded interval $I\subset\mathbb R$, let $\mathfrak A(I)$ be the $C^*$-algebra generated by Weyl elements whose test functions have support in $I$. The corresponding local von Neumann algebra is
	\begin{equation}
		\mathcal M_\omega(I) =\pi_\omega\big(\mathfrak A(I)\big)''.
	\end{equation}
	
	Local observable algebras in continuum quantum field theory are typically of type III. In particular, the interval algebras of the standard vacuum current net are type $\mathrm{III}_1$ factors. This local structure is distinct from the global algebra in the irreducible vacuum representation, which is $\mathscr{B}(\mathcal H_0)$ and hence of type I. Related constructions have found applications in AdS black holes, JT gravity, and cosmology, including de Sitter, inflationary, and centaur geometries. These developments use modular theory and gravitational constraints, often through crossed-product constructions, to obtain observable algebras whose entropy is related to generalized entropy in specified semiclassical regimes \cite{Leutheusser:2021qhd,
		LeutheusserLiu:EmergentTimes, LeutheusserLiu:SubregionSubalgebra, Witten:2021unn, Chandrasekaran:LargeNEntropy, Penington:2023dql,
		Chandrasekaran:DeSitterAlgebra, Chen:2024rpx, KudlerFlam:ObservationalCosmology, AguilarGutierrez:CentaurAlgebra, Espindola:2026uqa, Jensen:GeneralSubregions}.
	
	For the Weyl algebra with a Gaussian characteristic function, the covariance $S$ and the symplectic form $\sigma$ jointly determine the one-particle structure underlying the GNS representation. For the pure vacuum state constructed below, this yields the usual bosonic Fock representation. Mixed quasi-free states, including thermal states, admit related Fock-space realizations, typically involving a doubled one-particle space through the Araki–Woods construction.

	%%%%%%%%%%%%%%%%%%%%%%%%%%%%%%%%%%%%%%%%%%%%%%
	\subsection{Definition of the vacuum via positive energy}
	%%%%%%%%%%%%%%%%%%%%%%%%%%%%%%%%%%%%%%%%%%%%%%
	We now specialize to the Dirichlet-like boundary conditions introduced in Sec.~\ref{Dirichlet-BC}. For fixed lapse and shift, the conserved boundary energy is given by $\mathscr{E}$. Accordingly, we identify $\hat H \equiv \, :\!\hat{\mathscr{E}}\!:$ as the Hamiltonian generating normalized boundary time translations, with the normalized time coordinate defined by $\tau = N_\circ t$. We impose $N_\circ \geq |N_\circ^x|$, ensuring that the normalization of the boundary time evolution is compatible with the prescribed lapse and shift.

	We choose a centered quasi-free vacuum state $\omega_0$ invariant under this evolution and satisfying the \emph{spectrum condition}. More precisely, in its GNS representation $(\pi_0,\mathcal{H}_0,\Omega_0)$, time translations are unitarily implemented by a strongly continuous one-parameter group whose generator satisfies
	\begin{subequations}
		\begin{align}
			& \hat{H} \Omega_0=0\, , \\
			& \hat{H}\ge 0\, ,\\
			& \pi_0(\alpha_\tau(\hat{A})) = e^{i\hat{H}\tau}\pi_0(\hat{A})e^{-i\hat{H}\tau}\, .
		\end{align}
	\end{subequations}
	Here $\alpha_\tau$ denotes the time-translation automorphism of the algebra $\mathfrak{A}$ induced by the equations of motion. The first relation fixes the vacuum energy to zero and ensures that the vacuum vector is invariant under time translations. The second requires the energy spectrum to lie in $[0,\infty)$. The third states that the unitary group generated by the energy implements translations on the observables in the GNS representation.
	
	To determine the induced action on the smearing functions at fixed time, recall from Sec.~\ref{Dirichlet-BC} that the characteristic coordinates are $z^\pm=\tilde{x}\pm\tau$, with $\tilde{x}=x+N_{\circ}^x\, t$. The chiral currents therefore evolve according to $\hat{\mathcal J}_\pm(\tau,\tilde{x})=\hat{\mathcal J}_\pm(0,\tilde{x}\pm\tau)$. Changing the integration variable in the smeared currents then yields the induced translations of the smearing functions
	\footnote{In particular, changing the integration variable gives
		\begin{equation*}
			\int_{\Sigma}\dif \tilde{x}\,\eta_\pm(\tilde{x})\hat{\mathcal J}_\pm(0,\tilde{x}\pm\tau)
			=
			\int_{\Sigma} \dif \tilde{x}\,\eta_\pm(\tilde{x}\mp\tau)\hat{\mathcal J}_\pm(0,\tilde{x})\, .
		\end{equation*}
	}
	\begin{equation}
		(U_\tau\eta_\pm)(\tilde{x})
		=
		\eta_\pm(\tilde{x}\mp \tau),
		\quad
		\alpha_\tau\big(\hat W(\epsilon)\big)
		=
		\hat W(U_\tau\epsilon).
	\end{equation}
	Consequently, the one-parameter family of automorphisms $\alpha_\tau$ acts on the Weyl algebra by translating the smearing functions through the maps $U_\tau$. The opposite signs in these translations encode the opposite propagation directions of the two chiral currents. Thus, the time evolution of the algebra is implemented through the transport of its smearing functions, rather than through any change in the defining form of the Weyl operators.

	%%%%%%%%%%%%%%%%%%%%%%%%%%%%%%%%%%%%%%%
	\subsection{Continuum mode decomposition}
	%%%%%%%%%%%%%%%%%%%%%%%%%%%%%%%%%%%%%%%
	
	On the noncompact spatial slice $\Sigma$, we expand a test function at fixed time using the Fourier convention 
	\begin{equation}
		\eta_{\pm}(z^\pm) = \frac{1}{2\pi} \int_{-\infty}^{+\infty} \dif p \, \eta_p^\pm\, e^{- ipz^\pm} \, ,
	\end{equation}
	where reality of $\eta_\pm$ requires
	\begin{equation}
		\overline{\eta_p^\pm}=\eta_{-p}^\pm \, .
	\end{equation}
	We may evaluate these expressions on the slice $\tau=0$, where $z^\pm=\tilde{x}$. The corresponding smeared current can then be expressed in terms of the Fourier modes as
	\begin{equation}
		\hat{\mathrm J}_\pm(\eta_\pm) =   \frac{1}{2\pi}  \int_{-\infty}^{+\infty} \dif p \, \eta_p^\pm\,  \hat{\mathrm J}_p^\pm\, ,
	\end{equation}
	where
	\begin{equation}
		\hat{\mathrm J}_p^\pm :=\int_{\Sigma} \dif z^\pm \,\hat{\mathcal{J}}_\pm(z^\pm) \, e^{- ipz^\pm} \, .
	\end{equation}
	In contrast to the discrete modes encountered on a compact spatial slice, the Fourier modes on $\Sigma$ form an operator-valued distribution labeled by a continuous real parameter. The oscillator sector may be parametrized by $p\in\mathbb R_+$, while the global currents require a separate prescription. The continuum modes satisfy
	\begin{equation}
		[ \hat{\mathrm J}_p^\pm, \hat{\mathrm J}_q^\pm ] =\mp 4 \pi p\, \delta (p+q) \, , 
	\end{equation}
	and Hermiticity of the currents gives $(\hat{\mathrm J}_p^\pm)^\dagger=\hat{\mathrm J}_{-p}^\pm$. The opposite signs in the two chiral sectors can be accommodated by interchanging the roles of positive and negative momentum when defining annihilation and creation operators. In particular, we introduce
	\begin{equation}
		\hat{\mathrm{a}}_p^\pm  =  \frac{1}{\sqrt{4 \pi p}}\, \hat{\mathrm{J}}_{\mp p}^\pm \, , \qquad (\hat{\mathrm{a}}_p^\pm)^\dagger = \frac{1}{\sqrt{4 \pi p}}\, \hat{\mathrm{J}}_{\pm p}^\pm \, ,
	\end{equation}
	for all $p \in \mathbb{R_+}$. With these conventions, both chiral sectors acquire the same canonical continuum normalization,
	\begin{equation}
		[\hat{\mathrm{a}}_p^\pm, (\hat{\mathrm{a}}_q^\pm)^\dagger] =  \delta (p-q) \, , \qquad\text{for all } p>0\, .
	\end{equation}
	Here the operators labelled by $p>0$ should be understood in the usual
	distributional sense: individual $\hat{\mathrm{a}}_p^\pm$ are not, in
	general, operators associated with normalizable one-particle states, whereas their smeared versions define the corresponding operators on the Fock space.

	\begin{definition}
		The GNS vector $\Omega_0$ associated with the vacuum state $\omega_0$ is characterized, up to a phase, by
		\begin{equation}
			\hat{\mathrm{a}}_p^{\pm}\,\Omega_0 = 0 \qquad\text{for all~} p>0\, ,
		\end{equation}
		understood distributionally, meaning that every annihilation operator smeared with a square-integrable wave packet annihilates $\Omega_0$. The corresponding GNS Hilbert space is the tensor product of two bosonic Fock spaces,
		\begin{equation*}
			\mathcal{H}_0 = \bigoplus_{N=0}^\infty \operatorname{Sym}^N\!\left(L^2(\mathbb R_{+},\dif p)\right)
			\otimes
			\bigoplus_{N=0}^\infty \operatorname{Sym}^N\!\left(L^2(\mathbb R_{+},\dif p)\right) ,
		\end{equation*}
		corresponding to the chiral and antichiral sectors. The finite-particle vectors obtained from $\Omega_0$ by applying creation operators $(\hat{\mathrm a}_p^\pm)^\dagger$ smeared with square-integrable wave packets span a dense subspace of $\mathcal H_0$.
	\end{definition}
	
	In this Fock realization, the one-particle phase $e^{ip\tau}$ shows that a mode with momentum magnitude $p>0$ has energy $p$ in either chiral sector. Translations in normalized boundary time are implemented by the self-adjoint Hamiltonian
	\begin{equation}
		\hat H=\dif \Gamma(p\oplus p),
	\end{equation}
	where $p$ acts by multiplication on each one-particle space and $\dif \Gamma$ denotes bosonic second quantization. Equivalently, in the oscillator notation,
	\begin{equation}
		\hat H
		=\int_0^\infty \dif p\,p\,
		\Bigl( (\hat{\mathrm a}_p^+)^\dagger\hat{\mathrm a}_p^+ +(\hat{\mathrm a}_p^-)^\dagger\hat{\mathrm a}_p^- \Bigr)\, .
	\end{equation}
	
	The individual operators at a fixed momentum should be understood in the usual distributional sense, whereas smeared creation operators define genuine operators on the appropriate dense domain of Fock space.
	
	A direct computation yields
	\begin{equation}
		\langle\Omega_0,\hat Q(\epsilon_1)\hat Q(\epsilon_2)\Omega_0\rangle
		=S_0(\epsilon_1,\epsilon_2)+\frac{i}{2}\sigma(\epsilon_1,\epsilon_2).
	\end{equation}
	Here, the symmetric vacuum covariance is given by
	\begin{equation}
		\hspace{-0.3 cm}
		S_0(\epsilon_1,\epsilon_2)
		=\frac{1}{\pi}\int_{0}^{+\infty}\dif p\,p\,
		\Re\Bigl\{
		\overline{\eta_{1,p}^+}\,\eta_{2,p}^+
		+\eta_{1,p}^-\,\overline{\eta_{2,p}^-}
		\Bigr\}.
	\end{equation}
	For the test functions introduced in Sec.~\ref{sec:XVA}, the integrals converge, and the normalization is consistent with the oscillator commutation relations given above.
	
	The real and imaginary parts of the vacuum two-point function thus reproduce, respectively, the symmetric covariance and one-half of the symplectic form, with the latter fixed by the commutator of the smeared observables. The covariance satisfies the required positivity inequality, as follows from the Cauchy--Schwarz inequality in the underlying one-particle Hilbert space. Moreover, the vacuum expectation values of the corresponding Fock-space Weyl operators define a Gaussian characteristic functional. This establishes the positivity and regularity of the associated quasi-free vacuum state.
	
	The vacuum covariance is invariant under translations,
	$S_0(U_\tau\epsilon_1,U_\tau\epsilon_2)=S_0(\epsilon_1,\epsilon_2)$.
	Consequently, the corresponding vacuum state is translation invariant,
	$\omega_0\circ\alpha_\tau=\omega_0$.

	The global charge associated with the spatially integrated current requires separate consideration. It is formally written as 
	\begin{equation}
		\hat{\mathrm J}_0^\pm
		=
		\int_{\Sigma}dx\,\hat{\mathcal J}_\pm(x) \, ,
	\end{equation}
	whose operator meaning requires a limiting prescription in the chosen representation, since the constant smearing function does not belong to $\mathbb T(\Sigma)$.
	
	This distinction is essential on $\Sigma=\mathbb R$: the zero mode is a global charge associated with the spatially integrated current, whereas the operators with $p>0$ describe the continuum of current
	fluctuations. Moreover, since the modes are operator-valued distributions, the endpoint $p=0$ cannot in general be identified with an ordinary member
	of the continuum without specifying an appropriate limiting prescription. Classically, the charge is central relative to the current algebra when its Poisson brackets with the currents vanish under the adopted boundary conditions. Quantum mechanically, centrality additionally requires a well-defined charge whose unitary group commutes with the represented current algebra. Thus, on a noncompact slice, the global zero mode must be distinguished from the continuum of nonzero-momentum modes, in contrast with the compact periodic case, where constant smearing is admissible and the zero mode appears explicitly in the discrete Fourier decomposition.

	The momentum-labelled one-particle states
	$|p\rangle_\pm := (\hat{\mathrm{a}}_p^{\pm})^\dagger \Omega_0$
	are therefore not elements of $\mathcal{H}_0$ for fixed $p$. Rather, they are generalized momentum eigenstates with distributional normalization
	$\langle p|p'\rangle=\delta(p-p')$. The Dirac delta is a distribution rather than an ordinary finite inner product, so these states do not possess a finite Hilbert-space norm and cannot be used as ordinary vectors in $\mathcal{H}_0$. They are instead naturally interpreted within a rigged Hilbert space (Gelfand triple),
	\begin{equation}
		\varphi \;\subset\; \mathcal{H}_0 \;\subset\; \varphi^\times\, ,
	\end{equation}
	where $\varphi$ is a suitable dense nuclear subspace of sufficiently regular vectors and $\varphi^\times$ denotes its continuous antilinear dual. In this framework, the generalized eigenstates $|p\rangle_\pm$ are continuous antilinear functionals on $\varphi$, while ordinary wave-packet states obtained by smearing them with functions in $L^2(\mathbb R_+,\dif p)$ belong to the Hilbert space. Thus, the appearance of continuum-labelled momentum eigenstates does not signal additional Hilbert-space vectors; it reflects the continuous spectral decomposition associated with the noncompact spatial direction.

	%%%%%%%%%%%%%%%%%%%%%%%%%%%%%%%%%%%%%%%%%%%%%
	\section{\texorpdfstring{$\sqrt{T\bar T}$}{sqrtTTbar} radial flow}
	\label{sqrtTTbar radial flow}
	%%%%%%%%%%%%%%%%%%%%%%%%%%%%%%%%%%%%%%%%%%%%%
	
	The fluid/gravity correspondence developed in Secs.~\ref{sec:fluid-gravity}--\ref{sec:Boundary-metric} identifies the gravitational phase-space variables with the hydrodynamic fields of a two-dimensional fluid. Equation \eqref{eq:fluid-density-relations} characterizes the fluid in the conformal representative specified at $r=1$. To investigate the radial evolution of the fluid, we extend this identification to constant-$r$ hypersurfaces and determine the corresponding stress-tensor components as functions of the radial coordinate. This analysis yields an exact classical flow equation of square-root form, associated with the $\sqrt{T\bar T}$ deformation~\cite{Rodriguez:2021tcz}. We subsequently show that the radial evolution can be integrated into a finite $SO(1,1)$ transformation.
	
	Throughout this section, radial derivatives are taken at fixed $t$, $x$, and gravitational phase-space functions $(\mathcal P,\mathcal L)$. The chemical potentials are also held fixed under radial evolution. Consequently, the radial dependence of the fluid variables arises from the explicit dependence of the induced conformal metric on $r$, rather than from radial evolution of the underlying phase-space fields or chemical potentials.
	
	%%%%%%%%%%%%%%%%%%%%%%%%%%%%%%%%%%%%%%%%%%%%%
	\subsection{Fluid variables on constant-$r$ hypersurfaces}
	%%%%%%%%%%%%%%%%%%%%%%%%%%%%%%%%%%%%%%%%%%%%%
	
	We begin by extracting the fluid variables from the conformal metric $\sigma_{ab}$ defined in \eqref{Uni-Mod-Metric}. The gravitational fields $\mathcal P$ and $\mathcal L$, together with their associated chemical potentials, are independent of the radial coordinate. Expressing $\sigma_{ab}$ in ADM form, we identify the lapse and shift as
	\begin{subequations}
		\begin{align}
			N(r)&=\frac{4r \left(\mu\mathcal P-\nu\mathcal L\right)}
			{\mathcal P^2-r^2\mathcal L^2}\,,\\
			N^x(r)&=\frac{4 \left(\nu\mathcal P-r^2\mu\mathcal L \right)}{\mathcal P^2-r^2\mathcal L^2}\,.
			\label{eq:root-flat-adm-functions-new} 
		\end{align}
	\end{subequations}
	These expressions encode the radial dependence of the ADM decomposition while keeping the gravitational phase-space data and chemical potentials fixed.
	
	To determine the corresponding fluid variables, we evaluate the renormalized symplectic potential in the form \eqref{SP-Hydro}, replacing the boundary metric $\gamma_{ab}$ with $\sigma_{ab}$. In performing this identification, we retain the field-space variations of both the gravitational fields and the chemical potentials. The resulting energy density and momentum density are
	\begin{subequations}\label{rho-J-r}
		\begin{align}
			\rho(r) &=\frac{1}{8r}\left(
			\mathcal P^2+r^2\mathcal L^2\right),
			\\
			J &=-\frac14\mathcal P\mathcal L\,,
		\end{align}
	\end{subequations}
	with the pressure satisfying $p(r)=\rho(r)$. The energy density therefore acquires a nontrivial radial dependence, whereas the momentum density remains constant along the radial evolution. In particular, the latter is determined entirely by the gravitational phase-space fields and is insensitive to the explicit radial dependence of the conformal metric.
	
	At $r=1$, the expressions above reduce to the fluid-density equations in \eqref{eq:fluid-density-relations}, thereby recovering the stress-tensor variables associated with the chosen conformal representative. This provides the reference configuration from which the radial flow can be studied. In the following, we use these radial fluid variables to derive the square-root flow equation and its finite $SO(1,1)$ representation.
	
	%%%%%%%%%%%%%%%%%%%%%%%%%%%%%%%%%%%%%%%%%%%%%%%%%
	\subsection{The square-root radial equation}
	%%%%%%%%%%%%%%%%%%%%%%%%%%%%%%%%%%%%%%%%%%%%%%%%%
	
	In an orthonormal frame adapted to the induced conformal metric $\sigma_{ab}$, the covariant components of the stress tensor take the form
	\begin{equation}
		T_{\text{\tiny (frame)}}=
		\begin{pmatrix}
			T_{uu} & T_{us}\\
			T_{us} & T_{ss}
		\end{pmatrix}.
	\end{equation}
	We introduce the square-root stress-tensor invariant
	\begin{equation}
		\mathcal{R}(r):=\sqrt{\det T_{\text{\tiny (frame)}}}\, ,
		\label{eq:root-flat-operator-definition-new}
	\end{equation}
	where the positive branch is understood whenever the determinant is nonnegative.
	\footnote{The square-root operator is defined by
		\begin{equation*}
			\mathcal{R}(r)\equiv
			\mathcal O_{\sqrt{T\bar T}}(r)
			=\sqrt{\frac12\mathcal T^{ab}\mathcal T_{ab}
				-\frac14(\mathcal T^a{}_a)^2}\,,
		\end{equation*}
		where indices are raised and lowered using $\sigma^{ab}$ and $\sigma_{ab}$. The equality with the determinant expression follows from the two-dimensional algebraic identity for a symmetric rank-two tensor, with the Lorentzian signature understood. In our case, $\mathcal{T}^{ab}$ is traceless.}
	The local energy-dominant condition $\rho(r)\geq |J|$ ensures that the radicand is nonnegative for the stress-tensor parametrization used here, so that $\mathcal{R}(r)$ is real.
	
	To describe the radial evolution, we choose $r=1$ as a reference hypersurface and denote the corresponding fluid variables by $\rho(1)$, $\mathcal{R}(1)$, and $J(1)$. The logarithmic radial coordinate $\beta=\ln r$ measures the displacement from this reference surface, with $\beta=0$ at $r=1$. Using \eqref{rho-J-r} and \eqref{eq:root-flat-operator-definition-new}, the radial evolution equations become
	\begin{subequations}\label{eq:root-flat-linear-system-new}
		\begin{align}
			& \partial_\beta\rho(\beta)=-\mathcal R(\beta)\,, \qquad
			\partial_\beta\mathcal R(\beta)=-\rho(\beta)\,, \\
			& \partial_\beta J(\beta)=0 \, .
		\end{align}
	\end{subequations}
	The energy density and the square-root invariant therefore form a closed linear system, whereas the momentum density remains constant along the radial flow. In particular, the evolution preserves the difference $\rho^2-\mathcal{R}^2$, which is a useful consistency check on the flow and constrains the allowed radial dependence of these observables.
	
	%%%%%%%%%%%%%%%%%%%%%%%%%%%%%%%%%%%%%%%%%%%%%%%%%%%%
	\subsection{Exact integration and membrane interpretation}
	%%%%%%%%%%%%%%%%%%%%%%%%%%%%%%%%%%%%%%%%%%%%%%%%%%%%
	
	The system \eqref{eq:root-flat-linear-system-new} can be integrated exactly. Expressing the solution in terms of the reference data gives
	\begin{align}
		\begin{pmatrix}
			\rho(r)\\
			\mathcal R(r)
		\end{pmatrix}
		&=
		\begin{pmatrix}
			\cosh\beta& -\sinh\beta\\
			-\sinh\beta& \cosh\beta
		\end{pmatrix}
		\begin{pmatrix}
			\rho(1)\\
			\mathcal R(1)
		\end{pmatrix}\,.
		\label{eq:root-flat-finite-boost}
	\end{align}
	The radial evolution thus acts on $(\rho,\mathcal{R})$ as a finite $SO(1,1)$ boost, with rapidity $\beta=\ln r$ (see \cite{Bagchi:2022nvj,Bagchi:2024unl,Banerjee:2026qyc} for relevant discussion). This representation makes the composition law of successive radial displacements explicit: two admissible displacements combine by adding their logarithmic radial parameters. The inverse transformation is obtained by reversing the radial displacement.
	
	The boost structure also clarifies which features of the fluid variables are invariant under radial evolution. In addition to the constancy of $J$, as said before, the quadratic combination $\rho^2-\mathcal{R}^2$ is preserved. Consequently, the radial flow changes the individual values of the energy density and the square-root invariant while maintaining this quadratic relation. The reality condition on $\mathcal{R}$ must, however, continue to hold throughout the interval under consideration; the algebraic solution alone does not guarantee that every point along an arbitrary radial trajectory defines an admissible timelike hypersurface.
	
	From the membrane-paradigm perspective~\cite{Iqbal:2008by}, this result provides an explicit finite radial map between quasilocal fluid observables evaluated on different hypersurfaces, with the underlying gravitational configuration held fixed. Its interpretation as a relation between physical membrane variables requires that the hypersurfaces remain timelike, the spatial cuts remain spacelike, and the conformal normalization used to define the stress tensor remain regular. Subject to these conditions, the evolution is determined entirely by the radial separation and the initial fluid variables, without requiring a perturbative expansion in $\beta$.
	
	The construction is local on the boundary and does not require the spatial slice to be compact. For noncompact slices, or for compact slices with nontrivial periodicity conditions, the passage from local fluid variables to integrated charges additionally requires suitable endpoint or periodicity conditions and the convergence of the relevant integrals.

	%%%%%%%%%%%%%%%%%%%%%%%%%%%%%%%%%%%%%%%%%%%%%%
	\section{Discussion}\label{sec:conc}
	%%%%%%%%%%%%%%%%%%%%%%%%%%%%%%%%%%%%%%%%%%%%%%

	\paragraph{Summary of results.}
	We have formulated three-dimensional flat gravity as an
	$\mathfrak{isl}(2,\mathbb{R})$ Chern--Simons theory and studied the restricted
	phase space whose boundary data are the Cartan components $(\mathcal L,\mathcal P)$
	and their chemical potentials $(\nu,\mu)$. The charge algebra is an infinite-dimensional
	Heisenberg algebra, which diagonalizes into two commuting chiral current algebras
	$\mathcal J_\pm$ with opposite-sign central terms. Through the fluid/gravity dictionary
	of Sec.~\ref{sec:fluid-gravity}, these variables determine the stress tensor of a
	conformal fluid with $p=\rho$ on the boundary metric $\gamma_{ab}$.
	Regularity of the Weyl rescaling at $r=1$ yields the dominant energy bound
	$\mathscr E\geq|\mathscr J|$.
	
	For a non-compact spatial slice, finiteness of the zero modes, energy and momentum
	selects the class $\mathcal J_\pm\in L^1(\mathbb R)\cap L^2(\mathbb R)$. In place of
	a monodromy matrix, each chiral current defines a one-dimensional Schr\"odinger
	operator, and the global data are carried by the transport matrix relating the
	asymptotic solutions at the two ends of the line (Sec.~\ref{app:schrodinger-transport-matrix}).
	We considered two choices of boundary conditions.
	For the Heisenberg-like conditions, the chemical potentials are constant and the
	currents are stationary. Integrability of the symplectic form together with
	vanishing of the symplectic potential makes $\mathscr F(\mathrm P_0,\mathrm L_0)$, introduced in \eqref{cal-F-def},
	homogeneous of degree two and gives a first-law relation.
	For the Dirichlet-like conditions, the two chiral sectors propagate independently with
	velocities $v_\pm=N_\circ\pm N^x_\circ$. The chiral fields support travelling
	kinks of $\Phi_\pm$, whose jump is fixed by the zero mode $\mathrm J_0^\pm$,
	while the currents are localized, e.g. $\sech^2$ pulses \eqref{J-Sech-ex}. The associated
	Schr\"odinger operator for the given example \eqref{J-Sech-ex} is the attractive P\"oschl--Teller operator, which is reflectionless for integer $l_\pm$. Here the energy and momentum are conserved, and the
	homogeneity condition makes entropy $\mathscr S(\mathscr E,\mathscr J)$ homogeneous of degree one.
	We also set up the quantization on the real line as a Weyl $C^*$-algebra with
	continuous momentum labels, quasi-free vacuum and thermal states, and the associated GNS
	representation.
	
	Finally, keeping the radial dependence of the induced conformal metric, we found
	that the quasi-local stress tensor obeys an exact classical flow,
	$-r\partial_r\rho=\mathcal R$ on the fluid branch, with $\partial_r J=0$.
	This is the flat-space counterpart of the square-root operator
	associated with $\sqrt{T\bar T}$, and it integrates to a finite $SO(1,1)$ boost
	of $(\rho,\mathcal R)$ with rapidity $\beta=\ln r$. In contrast to the trace
	relations and Fefferman--Graham expansions used in the AdS$_3$ finite-cutoff
	literature, the flow is local, non-perturbative, and closed in the stress tensor alone.

	%%%%%%%%%%%%%%%%%%%%%%%%%%%%%%%%%%%%%%%%%%%%%%
	\paragraph{Membrane paradigm.}
	%%%%%%%%%%%%%%%%%%%%%%%%%%%%%%%%%%%%%%%%%%%%%%
	
	The membrane paradigm trades the null generators of a horizon for a
	fictitious, dissipative fluid membrane living on a nearby \emph{timelike}
	surface, the stretched horizon, on which physical quantities remain
	finite and observer-dependent redshift effects can be tracked
	explicitly \cite{Damour:1979wya,Thorne:1986iy,Hubeny:2010ry}. The construction of
	Secs.~\ref{sec:fluid-gravity}--\ref{sec:Boundary-metric} furnishes an
	explicit, fully covariant realization of this idea for asymptotically
	locally flat backgrounds, in line with the fluid/gravity perspective. The radial flow of Sec.~\ref{sqrtTTbar radial flow} supplies the corresponding
	radial evolution law: at fixed gravitational data, changing the membrane position
	is an $SO(1,1)$ transformation of $(\rho,\mathcal R)$ with the momentum density
	unchanged, in the spirit of the radial viewpoint of \cite{Iqbal:2008by}.
	
	\paragraph{True horizon versus stretched horizon.}
	
	The induced metric on a constant-$r$ hypersurface \eqref{eq:boundary-metric}
	is Lorentzian for any $r>0$ in the non-degenerate sector
	$\mu\mathcal P-\nu\mathcal L\neq0$. It degenerates precisely at $r=0$, the
	Rindler horizon. In the terminology of the membrane paradigm, the true
	horizon sits at $r=0$, while \emph{any} surface at finite $r=r_s>0$
	qualifies as a stretched horizon: a regulator surface on which the
	degenerate null geometry is replaced by a genuine Lorentzian one, and on
	which the fluid stress tensor $\mathcal{T}^{ab}$ of \eqref{BY-EMT} is finite and well defined. The limit $r_s\to 0$
	recovers the true horizon, where the membrane description is expected
	to break down in the usual way (infinite local temperature, vanishing
	proper thickness).
	This is visible in the radial data \eqref{rho-J-r}: as $r\to0$,
	$\rho\simeq\mathcal P^2/(8r)$ diverges, while $J$ stays fixed.

	Because of \eqref{gamma-sigma}, the choice made in
	Sec.~\ref{sec:Boundary-metric}, the fluid variables $(\rho,p,J)$
	computed from~\eqref{SP-Hydro} are, by construction, precisely the
	components of the stress tensor of the membrane located at the
	stretched horizon $r_s=1$. This is not merely a normalization
	convenience: on the Rindler background~\eqref{eq:Rindler-metric}, a
	static observer at fixed $r$ moving along the orbits of $\partial_T$
	has proper acceleration $1/r$ and measures a local (Unruh)
	temperature $T_{\text{\tiny loc}} = 1/(2\pi r)$, which obeys the Tolman law
	$T_{\text{\tiny loc}}\, r = \text{const.}$
	relative to the surface gravity $\kappa=1$ of the horizon at $r=0$ in
	these coordinates.
	In this normalization, the surface $r=1$ is the locus at which the locally
	measured temperature coincides with the intrinsic horizon temperature
	$\kappa/2\pi$ of the degenerate surface at $r=0$.
	
	\paragraph{Beyond the stretched horizon.}
	Nothing in the derivation of Sec.~\ref{sec:fluid-gravity} singles out $r_s=1$ as
	special from the point of view of the underlying formalism. The fluid dictionary
	can be built on an \emph{arbitrary} constant-$r$ hypersurface: the induced metric
	\eqref{eq:boundary-metric}, the Weyl factor $W(r)$ and the rescaled conformal
	metric $\sigma_{ab}(r)$ are defined for generic $r$, and the identification
	\eqref{gamma-sigma} is a convenient normalization of an otherwise $r$-independent
	construction. The fluid variables $(\rho(r),p(r),J)$ of \eqref{rho-J-r} then
	describe the membrane at any radius, and the flow of
	Sec.~\ref{sqrtTTbar radial flow} relates them exactly.

	\paragraph{Carrollian versus timelike fluid descriptions.}
	
	The induced conformal metric degenerates at $r=0$ and as $r\to\infty$. This is the
	regime in which boundary physics is conventionally organized in Carrollian language,
	with a degenerate lightcone structure replacing the usual relativistic one
	\cite{Ruzziconi:2026bix,Hartong:2025jpp,Hartong:2026bvh,Ciambelli:2024swv}.
	Our construction shows that, on admissible timelike hypersurfaces, the boundary
	data can instead be packaged into the canonically conjugate pairs
	$(\mu,\mathcal P)$ and $(\nu,\mathcal L)$ (cf.\ \cite{Nekrasov:2025egx}), and that these
	yield an ordinary Lorentzian fluid description. The map between gravitational
	and hydrodynamic data is fixed by simultaneously equating (i) the gravitational
	equations of motion \eqref{eq:EOM-RPS} with the conservation law
	\eqref{Conservation-equations} on the background $\gamma_{ab}$, and (ii) the
	renormalized symplectic potential with the fluid variational formula
	\eqref{SP-Hydro}. This determines $\mathcal T^{ab}$, hence $\rho$, $p$ and $J$,
	in terms of $\mathcal L$ and $\mathcal P$.
	
	The symplectic pairing of the fields
	and the causal character of the induced metric are different questions: the
	fluid frame still requires $W>0$ and $\mu\mathcal P-\nu\mathcal L\neq0$, and a
	count of fields does not by itself impose these conditions or regularize the
	normalization at a null boundary. The paired variables therefore provide a
	Lorentzian fluid description on admissible timelike hypersurfaces, whereas a
	corresponding description of the null boundary itself would require a separate
	construction. We also note that the Carrollian geometry of the solution phase space is
	distinct from that of the spacetime null boundary \cite{Adami:2023wbe}. An odd or singleton number of boundary fields forces a genuinely Carrollian
	treatment of the degenerate solution phase space, while the even pairing realized
	here allows the null/degenerate boundary physics at $r=0$ and at
	null infinity to be reformulated entirely as ordinary timelike
	relativistic hydrodynamics.

	\paragraph{Non-compact boundaries and integrability.}
	On a non-compact slice there is no closed contour defining a monodromy, and the
	global data are carried by the transport matrix and scattering data of the
	chiral Schr\"odinger operators. The zero modes $\mathrm J_0^\pm$ are global
	labels of superselection sectors, finite thanks to $L^1\cap L^2$ falloff, and the
	conserved energy and momentum can be obtained from the transmission amplitude. An
	important open direction is a systematic treatment of integrability for
	Chern--Simons-like theories of gravity \cite{Bergshoeff:2014bia,Merbis:2014vja,Adami:2017phg} with non-compact boundaries. This includes
	the flat KdV hierarchy \cite{Fuentealba:2017omf,Oblak:2020bkg,Vitel:2026ssa}
	and its AdS$_3$ counterparts \cite{Cardenas:2025qka,Pino:2025crn}, for which
	scattering data would replace monodromy as the organizing structure. It also
	includes the bi-Hamiltonian structure of \cite{Adami:2025pfk,Adami:2026jmo}, which should
	translate to the present framework through the dictionary \eqref{dictionary} and
	which we expect to underlie the first-law relations found for the Heisenberg-like
	and Dirichlet-like sectors \cite{Gibbons:1976ue}. The one-soliton sector, with
	its thermodynamic and geometric properties, deserves a detailed analysis.
	
	\paragraph{Soft hair and P\"oschl--Teller eigenfunctions.}
	Under the Dirichlet-like conditions, the kink of $\Phi_\pm$ has a net change fixed
	by the zero mode, and the corresponding current defines the P\"oschl--Teller
	operator, which is reflectionless for integer $l_\pm$. The bound-state
	eigenfunctions of this operator are localized around the kink, and their relation
	to the soft-hair excitations of the boundary \cite{He:2014laa,Hawking:2016msc,Hawking:2016sgy,Afshar:2016wfy,Haco:2018ske}
	is an open question. A natural conjecture is that the discrete spectrum counts
	independent soft-hair sectors; establishing this requires relating the
	transport matrix to the surface-charge algebra on the non-compact slice.
	
	\paragraph{Quantization and related flows.}
	At the quantum level, the Weyl-algebraic construction is insensitive to the
	non-compactness of $\Sigma$, but its representations and thermal interpretation are
	not: KMS states exist algebraically without a trace-class Gibbs operator, and the
	quantum definition of the zero modes needs a separate prescription. It remains to be seen whether the classical radial flow can be promoted to a quantum
	flow, and how it relates to the finite-cutoff $T\bar T$ and $\sqrt{T\bar T}$
	proposals in AdS$_3$ \cite{McGough:2016lol,Hartman:2018tkw,Rodriguez:2021tcz}.
	
	\paragraph{Related deformations and integrable flows.}
	
	A further direction is to connect the present construction with the
	auxiliary-field formulation of integrable deformations, which generates
	families of integrable sigma models and their higher-spin generalizations
	\cite{Ferko:2024ali,Bielli:2024ach}, and also \cite{Fukushima:2026znr}. Since the radial flow
	derived here is an exact, closed evolution of the stress tensor
	and the chiral sectors are integrable, it would be interesting to
	determine whether it can be recast in such a deformation language.

	\paragraph{Function Spaces and Asymptotic Conditions.}
	
	On non-compact spatial slices, specifying the appropriate function space is essential for ensuring that gravitational charges and the associated fluid quantities are well defined. We work with dynamical fields in $L^1(\Sigma)\cap L^2(\Sigma)$, which provide the integrability and square-integrability conditions adopted in our construction. Schwartz regularity is a stronger assumption that defines a convenient subclass, particularly for constructing the transfer matrix, the non-compact analogue of the monodromy matrix, and relating the formalism to known physical solutions. Relaxing square-integrability may admit additional configurations, such as kink solutions, whose treatment requires accounting for possible source contributions at spatial infinities.
	
	\paragraph{Radial Deformations and Carrollian Limits.}
	
	The radial coordinate also parametrizes the deformation flow. We take the undeformed theory to be defined at $r=1$, while radial evolution probes different regimes, including the approach to the Rindler horizon and null infinity, where Carrollian physics becomes relevant. This provides a framework for investigating the flat-space analogue of $\sqrt{T\bar T}$ deformation flows.
	
	\paragraph{Variational Principles and Boundary Conditions.}
	
	The condition $\delta S=0$ specifies the variational principle adopted here and constrains the boundary ensemble, but does not by itself reduce the number of independent charges. Alternatively, one may allow an integrable but non-vanishing variation of the action, corresponding to a different ensemble related by a Legendre transformation on phase space. Relaxing the requirement of vanishing symplectic flux further opens the possibility of leaky boundary conditions \cite{Fiorucci:2021pha,Grumiller:2023ahv}, under which energy and other conserved quantities can flow across the boundary. Integrable $(1+1)$--dimensional systems with analytically tractable equations of motion may provide a useful setting in which to investigate these extensions.

	%%%%%%%%%%%%%%%%%%%%%%%%%%%%%%%%%%%%%%%%%%%%%%%%%%%%%%%%%%%%%%%
	\section*{Acknowledgement}
	We would like to thank Hossein Babaei-Aghbolagh, Jan de Boer, Daniel Grumiller, Marc Henneaux, Patricio Salgado-Rebolledo and Edgar Shaghoulian for discussions and comments on the manuscript. RE is supported by the Shuimu Tsinghua Scholar Program. The authors acknowledge the use of ChatGPT (OpenAI) and Claude AI for linguistic editing, improving exposition, and discussing certain technical aspects. The authors take full responsibility for the content of this manuscript.

	%%%%%%%%%%%%%%%%%%%%%%%%%%%%%%%%%%%%%%%%%%%%%%%%%%%%%%%%%%%%%%%%%%%

	%%%%%%%%%%%%%%%%%%%%%%%%%%%%%%%%%%%%%%%%%%%
	\bibliographystyle{fullsort.bst}
	\bibliography{reference}

	\clearpage
	
	%%%%%%%%%%%%%%%%%%%%%%%%%%%%%%%%%%%%%%%%%%%
	
\end{document}